# Geometrical Field Formulation of Thermomechanics in Rational Mechanics

Xiao Jianhua

Measurement Institute, Henan Polytechnic University, Jiaozuo, 454000, China

**Abstract:** In modern science, the thermo mechanics motion can be traced back to quantum motion in micro viewpoint. On the other hand, the thermo mechanics is definitely related with geometrical configuration motion (phase) in macro viewpoint. On this sense, the thermomechanics should be formulated by two kinds of motion: quantum motion and configuration motion. Its principle goal ought to be bridge the gap between atomic physics and engineering practice. In this research, the configuration motion is formulated by deformation geometrical field (motion transformation tensor). The quantum motion is formulated by the wave function of quantum state. Based on these two fields, the thermo stress is formulated as the coupling of quantum motion and configuration motion. Along this line, the entropy is interpreted and formulated according to thermodynamics rules. For scalar entropy, the traditional meaning of entropy is reserved. For infinitesimal configuration variation, the formulation is degenerated to the traditional elasticity deformation. For large random configuration deformation, the formulation is degenerated to the statistical physics methods. This research supplies a possible formulation to bridge the gap between the macro deformation and the micro quantum motion.

**Contents**

## 1. Introduction

Although the different sections of science focus on their individual fields, the kernel of them is the matter motion in general sense. The macro continuum is the most common matter form. So, establishing a general form of continuum mechanics to cover a broad range of matter motion is valuable. These efforts are never really abandoned, though it is failed again and again in absolute sense. There is no doubt that to establish an universal formulation of matter motion which can be applied in all scientific braches is in vain. However, gradually extending the current formulation to catch a much wider phenomenon fields are practical. Rational mechanics takes this direction as its target. This research belongs to this category.

In this paper, the thermo mechanics is studied. As a macro phenomenon, the temperature variation will cause stress (pressure) or configuration (volume) variation. The reverse is true also. Therefore, once the traditional deformation mechanics takes the temperature variation as a principle quantity, the problem will be complicated (although it is more approaching reality) or becomes un-closed problem in mathematic sense. On the other side, the micro motion viewpoints equipped with statistics methods explain the temperature as the motion of quantum. Therefore, the quantum mechanics is used to establish the thermo quantity. On conceptual sense, this is very successful. However, once evaluating this relation by deterministic equations, the difficulty becomes apparent. Hence, for actual engineering application, there is an urgent need to formulate a suitable theoretic frame to promote the combination of quantum mechanics and continuum mechanics. Such a theoretic frame should be widely applicable for physical chemistry, physical biology, and many others in modern industry.

From practical viewpoint, the so-called knowledge explosion and scientific documentation explosion have supplied a lot of phenomenon data which are waiting to be explained by a unified theoretic formulation. Furthermore, the internet technology makes the data collection from diverse sources become efficient and practical. Hence, reviewing the reported phenomenon to establish a unified theoretic formulation becomes practical. At least it is true comparing with the conditions several tens years ago.

Unfortunately, after the failure of 1950s-1970s, there are little such efforts are published except in pure mathematics or theoretic physics fields. The results are that the continuum mechanics (which is more directly related with common phenomenon in industry) becomes very passive in modern industry. When the digital simulation becomes the hot point, the available theoretic formulations are used in their extreme. Experimental data are explained with bias or artificial skills. As a direct result, the gap between different scientific branches becomes deep and deep. The unity of matter motion is broken into pieces.

Luckily, although the diverse tendency is very apparent, the thermo mechanics does play the role to unify the matter motion in general sense. Hence, the key may be at handling the thermo mechanics in a rational form.

In modern science, the thermo mechanics motion can be traced back to quantum motion in micro viewpoint. Statistic physics plays the main role in this field. There, the basic matter element is particle or its similar concept. It is believed that the interaction among particles can be described by suitable correlation functions or state probability density functions. Although the possible density functions are limited by the allowable quantum motions, there is no definite evidence to prove that the matter motion does act like that. Therefore, to prove its effectiveness, digital simulation becomes the main topic along this line. By comparing with experimental results, it is true that many local features indeed can be well explained. However, it is difficult to relate these results with macro motion of matter as a whole.

Viewing this shortage, the thermo-mechanics along the traditional line is still active. Some efforts to formulate the thermo-mechanics as a tensor field had been done. However, failing to give out the exact geometrical meaning of thermo-quantity tensor makes these efforts baseless.

Based on micro viewpoint, the thermo mechanics is definitely related with geometrical configuration motion (phase). On this sense, the thermo mechanics should be formed by two kinds of motion: quantum motion and configuration motion.

As a basic effort, the entropy concept as a bridge between micro motion and macro motion is selected as the break-through-point for modern scientific researches. Recently, based on the locality viewpoint (individual constituents interact mainly with their few nearest neighbors), a good review is published (J Eisert, M Cramer, and M B Plenio, 2010, [1]). Where, the area laws for the entanglement entropy are reviewed. Reasoning from the distance dependence of correlation function will leads to the volume dependence of entanglement entropy. Which one we should believe? If one accepts the area laws, the surface dependence of entropy will suggest that the entanglement entropy quantity scaling is a boundary area variation feature of motion in geometrical sense rather than the distance variation feature. Reasoning from geometrical consideration, as the distance variation is related with covariant tensor bases, the logic conclusion of area laws is that the entanglement entropy quantity is related with anti-covariant tensor base. Reasoning from the implication of area laws on quantifying the effective degree of freedom in statistic physics, the motion related with entropy concept should be formulated by anti-covariant tensor bases, if a geometrical field theory be formulated for thermo mechanics. Hence, the logic solution is to take the basic material element as a distinguishable closed region. Then, the quantum motion relates the closed region and its exterior via the boundary surface. In fact, this model is already well used. The closed region may be a molecular or an atom. On this sense, it seems no essential changes with traditional particle models. However, the essential difference is that: the interaction among particles is via surface rather than the center distance. From macro motion points, the surface interaction is more closely related with macro phenomenon. Hence, the basic thermo mechanics should be formulated on the description of surface deformation of basic material elements. This forms the basic viewpoint of thermo motion in this research.

From macro viewpoint, relating the macro deformation of continuum with the general entropy by formulating the related macro equations in traditional thermo mechanics frame is not applicable in principle to deformed continuum. This topic is well reviewed by P W Bridgman (1950, [2]). As the macro deformation is related with stress while the entropy variation is related with temperature gradient, the kernel conclusion is that, except a common quantity is found to be the essential description of micro motion and macro motion, there is no way to establish the rational formulation for thermo mechanics. However, if one takes the area laws of entanglement entropy as a general rules, the area deformation concept indeed can take the role as the basic matter motion if the basic material element is defined as a closed region. Along this line, the configuration concept must be used to describe the matter motion in general.

On macro motion viewpoint, no matter what phase of continuum is, the pressure of the matter is universally existed for all physical real temperature. However, the configurations of matters are not universal. Roughly, the configurations of matter are classified as solid, liquid, and gas in traditional physics and chemistry. For modern physics, the condensed matter concept is introduced to represents the configuration concept. However, essential advancement is little. This only shows that the micro configuration and macro configuration are not related directly. Their relationship is not purely geometrical rather it is physically. Failing to distinguish this essential difference is the main cause for extending the macro deformation concept to micro scale or extending the local quantum motion to macro scale through correlation function.

Summering above reasoning, the macro configuration deformation related quantities (strain and stress) should be related with the temperature gradient which is related with micro material element surface deformation. The micro configuration deformation is related with temperature gradient. On this sense, the temperature is a global concept. As the matter units interaction happen with their neighbors, the local configuration concept is a locality features of matter motion. The distinguished interacting matter units set forms the entangle concept of matter. Based on this understanding, the area laws for the entanglement entropy are established with vigorous arguments (J. Eisert, et al., 2010, [1]). As the entanglement is area related, the correlation function failed to support the area laws. By practical points of views, the most common materials used in daily life have relative fixed configuration. Their thermo features variation indeed should be understood by the entanglement entropy rather than the correlation function concept. Following this concept line, the traditional thermo dynamics should based on entanglement entropy rather than the correlation function concept. This is the logic conclusion from the motion concept of relativity. Yes, for actual measured matter motion, the infinitesimal motion plays the real local motion role. Hence, this line forms the main stream of reasoning in this research.

Firstly, the basic material unit concept is described with a review about entropy concept. After that, the basic material element motion is expressed by its surface configuration variation. Based on this basic motion, macro deformation quantities are established. Then, the research goes directly to formulate the thermo mechanics in a rational form. The thermo mechanics is formed by two kinds of motion: quantum motion and configuration motion. In this research, the configuration motion is formulated by deformation geometrical field (motion transformation tensor). The quantum motion is formulated by the wave function of quantum state. Based on these two fields, the thermo stress is formulated as the coupling of quantum motion and configuration motion. Along this line, the entropy is interpreted and formulated. Theoretically, the entropy is a rank three tensor, related with the curvature tensor of motion space. For scalar entropy, the traditional meaning of entropy is reserved. For infinitesimal configuration variation, the formulation is degenerated to the traditional elasticity deformation. For large random configuration deformation, the formulation is degenerated to the statistical physics methods. This research supplies a possible formulation to bridge the gap between the macro deformation and the micro quantum motion.

## 2. Description of Material Element

How to define the basic material unit in thermo mechanics is a very important essential problem. In many books, this problem is simply replaced by a suitable assumption. In this research as the incremental motion or perturbation motion are discussed, a general unit material concept should be established by a much more general examination. For this purpose, the entropy variation concept is used as the basic quantity.

### 2.1 Simple Review of Entropy Concept to Establish the Basic Material Unit

In thermo-mechanics, the entropy concept plays a very important role. As a result, many arguments are published on this topic. The widely used entropy is defined as:

$$dS = \frac{dQ}{T} \tag{0-1}$$

Or in the integral form:

$$S_2 - S_1 = \int_{state1}^{state2} \frac{dQ}{T} \tag{0-2}$$

Their equivalence is vaguely proved in textbooks as a commonsense.

As the absolute temperature is defined by the intrinsic energy, or says the background energy, it is an universal physical quantity. Under this background, the entropy takes the role as "motion measure" in physical sense. Hence, the entropy is the measure of matter motion under an energy background expressed by temperature. Viewing the temperature as a scalar field, the entropy can be explained as the interaction between the field and matter under discussion. Hence, this topic is very general.

In statistic physics, the motion measure is expressed by the distribution of possible "states". As the distribution implies the matter points relationship, it is indeed is a measure of motion. In this research, the matter points-relations are treated by deformation geometrical theory. The main interesting is the entropy variation related with the configuration variation.

By reviewing different forms of the entropy function, W F Durand (1932, [3]) argued that the integration $\int dQ$ is path-dependent. Hence, the Equ.(0-1), true complete differential form of entropy, must be introduced. He pointed out that the temperature is independent with the detailed material and geometry. Hence, its role is similar with the mass concept in Newton mechanics. The logic conclusion is that the entropy is the basic representation of matter motion in abstract sense.

In state space $x$, for a given probability density function $f(x)$, the Boltzman-Gibbs entropy is defined as:

$$H(f) = -\int_X f(x) \ln f(x) \cdot dx \tag{0-3}$$

This definition makes the entropy be the quantity to be calculable from micro motion viewpoints. When the possible micro motion states cannot be automatically be reduced, the entropy cannot be reduced. Hence, for a closed system, maximum entropy principle is the logic results of such a kind of definition. M C Mackey (1989, [4]) gave out a good review on this topic.

No doubt, this reasoning is arguable. In fact, the entropy increasing is reasoning from Equ.(0-2) initially. W S Franklin (1910, [5]) argued that: entropy concept as a description of matter state changes makes it difficult to take it as an independent concept. Therefore, the entropy concept should be understood as the correspondence of time concept. Based on this understanding, the concept of entropy should be derived by applying the Hamilton principle to limit the possible states. If this concept is accepted, the increasing entropy leads to the 'time' direction irreversible.

For high mass density matter, under the same heat quantity input, the temperature increase is slower than less mass density matter. Reasoning along this direction, the entropy defined by Equs.(1) and (2) is mass dependent Then potential temperature $\theta$ concept is introduced to define the entropy per mass (L A Bauer, 1908, [6]) as:

$$S = C_P \cdot \ln\theta + const \tag{0-4}$$

This leads to the absolute temperature concept.

In statistic physics, the Boltzman entropy is widely used. It is defined as:

$$S = k_B \cdot \ln\Omega \tag{0-5}$$

The $\Omega$ is the state number density function. By this definition, the entropy is determined by state number density. The state number density is taken as logic independent quantity.

For infinitesimal motion of matter, the entropy is taken as an independent quantity and the state function is taken as the phenomenon observable. So, to find the state variation, the phenomenon quantity is defined as:

$$\Omega = \exp\left(\frac{S}{k_B}\right) \tag{0-6}$$

This forms the bases for the thermodynamic fluctuation theory. When the Riemannian geometry is used in this field, it was found that the density function is equivalent with the potential well function (G Ruppeinier, 1995, [7]). Through the potential function, the temperature variationis related with potential boundary. As the potential well determines the motion of quantum, the state space may be replaced by the quantum functions space. Hence, it is a way to approach the quantum mechanics. In such a formulation, the thermodynamic curvature is proportional to the inverse of the free energy and is produced by interaction. The important viewpoint of these researches is that the potential gradient plays an important role.

Once the potential gradient is introduced in thermodynamics as an essential role, the phase transition of matter can be related with the configuration space topology variations (Kastner, M., 2008, [8]). This is performed by the implied concept of temperature is equivalent with energy (hence, potential) concept which is widely accepted in modern physics.

For infinitesimal motion, such as the lattice vibrations, the entropy does have significant variations (A van del Walle and G Ceder, 2002, [9]). In this case, the partition function depends on the configuration-dependent energy. Reasoning along this line, the entropy is configuration dependent. In fact, in a review paper about 'liquid' (J A Barker, and D Henderson, 1976 [10]), the entropy concept is escaped by focusing the pressure-temperature relations. Its logic conclusion is that: the basic configuration, quantum potential well form, and partition function are interacting quantities. Returning to the area laws of entanglement entropy, a natural way is to study the entropy variation with the configuration variations.

To this target, the basic material element motion is expressed by its surface configuration variation. Therefore, the basic material unit is a finite body with closed boundary to form a closed surface. The basic motion is the deformation of the surface.

Based on the quantum mechanics viewpoint, such a boundary is defined by the gradient of potential function or says the well position of quantum potential. Generally speaking, when passing the boundary, the potential gradient will change its sign. Hence, the boundary position (well position) is defined by zero potential gradient position. For a given material element, the exterior action causes the change of the boundary configuration. As the potential well change will cause the corresponding change of quantum wave function, the quantum motion is related with the configuration deformation naturally.

**2.2 Geometrical and Physical Description of Unit Material**

Based on above reasoning, the basic material element geometrical motion is described by basic surface deformation. Hence, the basic matter unit is expressed by a local closed geometrical surface. For deformation geometry, Prof. Chen Zhida (1987, 2000, [11]) has established a finite deformation geometry theory for continuum. This research will use this geometry theory.

In three dimensional space, this local surface field or material basic unit is expressed by the basic area vector $\vec{g}_0^i, i = 1,2,3$. They form the basic anti-covariant vector bases. Therefore, the local interaction of surface fields (in continuum mechanics) is expressed the motion transformation tensor, defined by the following equation:

$$\vec{g}^i = G_j^i \vec{g}_0^j \tag{0-7}$$

Coordinating the continuum by anti-variant coordinators $x^i, i = 1,2,3$, then the motion transformation tensor is position dependent. As usual, the ideal basic material unit is in statistics sense. That is the basic material unit is idealized.

In fact, the lattice vibration can be expressed by this formulation when the standard rectangular coordinator system is used. In this special case, it can be approximated as a conformal field (van del Walle, 2002, [3]). The related experiments show that the lattice vibration has significant contribution to entropy. Based on this fact, the motion transformation is a practical quantity rather than artificial one, here.

As the motion transformation tensor is determined by the interaction matter with the outside-world, it is very general. Although the new local field $\vec{g}^i$ is a new local surface field viewed by mathematical sense, there is no way to conclude that the $G_j^i$ has conformal field features. Therefore, the motion description of conformal field theory is abandoned, here. In fact, if conformal field theory is used, the motion definition will be formulated as:

$$g^{ij} = \eta^2 \cdot g_0^{ij} \tag{0-8}$$

Its main shortage is omitting the local orientation variation and mixing the contribution from distance variation and surface variation.

In Chen rational mechanics, the motion transformation tensor is determined by the matter motion. On the other hand, to identity the same material units, the commoving dragging coordinators system are introduced to identify the locality of a material unit, no matter it is in its initial configuration or current configuration. (Note that, the material objectivity, the global translation invariant and global rigid rotation invariant are naturally reserved by the dragging features of coordinators.)

By this selection, the motion expressed by the motion transformation tensor is the real physical quantity which is looking for to express the interaction between material units set.

For plastic deformation, the entropy concept is examined (Bridgman, 1950, [2]) in macro viewpoint. In this research, more generalized deformation will be studied to extend the entropy concept in a rational way.

Physically, for unit material, its boundary is defined by the potential well. As the potential energy can be formulated by absolute temperature field, the directly measurable temperature gradient field will be used as the basic physical description of unit material. For a given temperature field $T(x,y,z,t) = T(x^1, x^2, x^3, t)$, it is scalar field. It is absolutely position dependent. (here, the temperature field term is used rather than the potential term, as they are changeable in quantum physics).

The temperature gradient can be constructed as:

$$\vec{q} = q_i \vec{g}^i = \frac{\partial T}{\partial x^i} \vec{g}^i \tag{0-9}$$

Here, the temperature gradient tensor is covariant. As the temperature is an universal measurable thermo quantity, it is selected as basic variables in this research. In this research, the time parameter is implied by the deformation concept where the initial configuration and current configuration concept are used for the comparing between two time points.

## 3. Temperature Gradient and Deformation Coupling

For thermo mechanics, the continuum may have no macro configuration described by distance variation may, but the micro unit surface variation related with temperature may exist, hence no macro displacement field does not means that there is no micro surface configuration deformation. In this research, the continuum is viewed by the unit surface deformation and the macro distance variation caused conventional deformation. They are coupled by the intrinsic features of the material. Hence, the volume variation can be expected. So, the thermo motion should be described by two kinds: micro surface deformation related with thermo motion and the macro distance deformation related with conventional deformation mechanics. To introducing their relationship, unit volume quantity is introduced as a measurable universal quantity. Hence, two thermo quantities are used as basic quantity: temperature and volume.

In Chen (1987, [11]) geometrical field theory deformation, for an objective material unit in dragging coordinator system, the surface deformation transformation is defined as:

$$\vec{g}^i = G^i_j \vec{g}^j_0 \tag{1}$$

The distance base vector transformation $F_i^{\ j}$ is related with initial basic vector by the following equations:

$$\vec{g}_i = F_i^{\,j} \vec{g}_j^{\,0} \tag{2-1}$$

Where, $\vec{g}_j^{\,0}$ is initial distance base vector, $\vec{g}_i$ is the current distance base vector. The form invariant volume condition of unit material in continuum will give out the equation:

$$F_i^l G_l^j = \frac{V}{V_0} \cdot \delta_i^j, \quad \vec{g}_0^{\,i} \cdot \vec{g}_j^{\,0} = V_0 \delta_j^i \tag{2-2}$$

Where, $V$ is current volume of unit material, $V_0$ is initial volume of the same unit material. In continuum thermo mechanics, this is an approximation. Generally speaking, for Riemann geometry theory where the matter configuration is invariant while the gauge tensors are arbitral selected, the equation (2-2) does not hold. However, for thermo motion in continuum geometry, the matter configuration is deformed. The surface deformation and the distance deformation are partially independent. Hence, the equation (2-2) simply means that the volume can be constructed by the two deformations. This equation cannot be justified by pure mathematics reasoning. Here, it can be viewed as a physical requirement. For arbitral deformation, the orthogonal feature cannot be maintained. However, as the unit material configuration in thermo mechanics is in a statistic sense, this equation should be understood as the independent features of surface deformation and distance deformation. The volume invariant condition should be viewed as an approximation in statistic sense.

For simplicity, the initial distance base vector and the dragging coordinator can be selected by the standard laboratory coordinator system. By such a selection, the conventional meaning of macro deformation is maintained.

In Riemannian geometry, $\frac{V}{V_0} = 1$ is taken as the basic condition for objective invariance of real configuration, where the basic material element has no physical deformation, the base vector transformation is caused by different coordinator system selections. However, for deformation geometry, the $V$ represents current unit material volume as the material unit is coordinated by fixed coordinators. Failing to recognize this point will cause serious miss-understanding. Generally, for deformed materials, it is not unit. This is the essential difference between Riemannian geometry and deformation geometry.

For simplicity, at the following formulation of this paper, the $V_0 = 1$ is taken as the initial unit volume. By this definition, the quantity $V$ will represent the current volume unit. Note that: for fixed unit center distance, the macro configuration (large amount of units) of continuum has no deformation. However, the unit volume variation caused by the unit surface deformation does exist. Therefore, the temperature gradient caused configuration deformation is different with the conventional distance variation caused macro configuration deformation. There relationship is expressed by equation (2-2).

**3.1 Temperature Gradient**

If the order sequence of molecular is fixed, but the molecular is not fixed (such as local orientation changes), the material will have deformation. For such a kind of deformation, the molecular symmetry will let the molecular can rotate partial-freely. However, the fixed order (sequence of molecular) still makes the macro material integrity. This is the kernel for temperature variations in continuum mechanics.

As a logic conclusion of the order sequence of molecular is fixed, in commoving dragging coordinator system, the scalar gradient $\frac{\partial T}{\partial x^i}$ is fixed. However, the physical component is gauge tensor dependent. That is:

$$\vec{q} = \frac{\partial T}{\partial x^i} \vec{g}^i = \sqrt{g^{(ii)}} \frac{\partial T}{\partial x^i} \cdot \frac{\vec{g}^i}{\sqrt{g^{(ii)}}} \tag{3}$$

By this feature, the surface deformation will cause the variation of macro-measurable-physical temperature gradient. In fact, one has:

$$\vec{q} = \frac{\partial T}{\partial x^i} \vec{g}^i = \frac{\partial T}{\partial x^i} G^i_j \vec{g}^j_0 \tag{4}$$

Comparing with the initial temperature variation vector:

$$\vec{q}_0 = \frac{\partial T}{\partial x^i} \vec{g}^i_0 \tag{5}$$

The temperature gradient variation vector of the same unit material can be defined as:

$$\delta \vec{q} = \frac{\partial T}{\partial x^i} \vec{g}^i - \frac{\partial T}{\partial x^i} \vec{g}^i_0 = \frac{\partial T}{\partial x^i} (G^i_j - \delta^i_j) \cdot \vec{g}^j_0 \tag{6}$$

As the strain can be defined by the deformation gradient, hence, the temperature variation can be related with stress, as:

$$G^i_j - \delta^i_j = D^{il}_{jk} \cdot \sigma^k_l \tag{7}$$

Therefore, once the physical components of temperature variation vector are measured as a known quantity, the scalar temperature gradient will be changed in standard laboratory coordinator system. That is:

$$\tilde{q}_i = \sqrt{g^{(ii)}} \frac{\partial T}{\partial x^i} \tag{8}$$

Hence, the scalar temperature gradient is obtained as:

$$\frac{\partial T}{\partial x^i} = \frac{\tilde{q}_i}{\sqrt{g^{(ii)}}} \tag{9}$$

Here, the gauge tensor is determined by unit surface deformation. So, it is a known quantity.

Then, the next step is to study the temperature gradient dependence on the micro motion in sequence order. This will come into quantum mechanics.

The reverse of the process will form a road from quantum mechanics to macro deformation. This is the target of "lattice dynamics".

In this formulation, if the temperature variation is defined as the energy transportation, then the temperature gradient is equivalent with the general surface force. Hence, when the area is defined as a closed surface, the average calculation of point number $(N_0)$ over the area is the key point for statistic physics.

In classical interpretation, as zero order approximation, the temperature increase is expressed as:

$$\Delta T = \int_0^l \frac{\tilde{q}_i}{\sqrt{g^{(ii)}}} dx^i = \frac{3 \cdot (E_0) \cdot (N_0)}{(A_0)} \tag{10}$$

For stochastic motion, the average total energy $3(E_0) \cdot (N_0)$ over average free area scale $(A_0)$ is determined as the temperature. The average calculation of point number $(N_0)$ over an area $(A_0)$ is interpreted as the average hitting number. So, an probability density function of hitting can be introduced. The detailed formulation along this line is available from statistic physics textbook. This line is abandoned in this research, as the statistic methods are applied too early to represent the real details of micro motion. The above explanation is for easy understanding.

**3.2 General Stress**

As a surface quantity, the temperature is equivalent with the potential function. By this understanding, the temperature variation vector is the surface force. Then, using equation (1), the heat stress should be understood as:

$$\frac{\partial \vec{q}}{\partial x^j} = \frac{\partial^2 T}{\partial x^i \partial x^j} \vec{g}^i + \frac{\partial T}{\partial x^i} \Gamma^i_{jl} \vec{g}^l = (\frac{\partial^2 T}{\partial x^i \partial x^j} + \frac{\partial T}{\partial x^k} \Gamma^k_{ji}) \vec{g}^i \qquad (11)$$

Hence, one has the general thermo stress definition:

$$\sigma_{ij} = (\frac{\partial^2 T}{\partial x^i \partial x^j} + \frac{\partial T}{\partial x^k} \Gamma^k_{ji}) \qquad (12)$$

For fixed scalar gradient (as an approximation of potential wall), the equation is simplified as:

$$\sigma_{ij} = \frac{\partial T}{\partial x^k} \Gamma^k_{ji} \qquad (13)$$

The only variable quantity is the link $\Gamma^k_{ij}$, which is completely determined by the current surface geometrical features. This way will lead to the statistical methods, where the different shape of material unit plays the key role. For isotropic stress (pressure) case, the link plays the role of probability density. Refer Born and Huang (1956, [12]).

In this research, the deformation tensor will be used. This is different from statistic physics.

By equation (4), the deformation form is obtained as:

$$\frac{\partial \vec{q}}{\partial x^j} = (\frac{\partial^2 T}{\partial x^l \partial x^j} G_i^l + \frac{\partial T}{\partial x^l} \frac{\partial G_i^l}{\partial x^j}) \vec{g}_0^i \qquad (14)$$

So, in rational mechanics, the general stress is defined as:

$$\sigma_{ij} = \frac{\partial^2 T}{\partial x^l \partial x^j} G_i^l + \frac{\partial T}{\partial x^l} \frac{\partial G_i^l}{\partial x^j} \qquad (15)$$

It shows that the stress tensor is composed by two parts. For continuum in thermo mechanics, the two order differential of temperature field can be approximated as zero for first order approximation. For traditional elastic mechanics, the unit surface deformation gradient can be omitted. Therefore, the first item on the right side is the intrinsic related item related with elastic stress.

For continuum mechanics point of view, when the temperature gradient is approximated by a delta function, the continuum is composed by isolated particles. In fact, the statistic energy density expressed by correlation distance plays such a role in statistic physics.

In micro scale, for periodic structure of potential field, the temperature scalar gradient is zero on the unit boundary surface. At the same time, it takes its maximum on the inner boundary surface and reverses its sign on the exterior surface. The balance of inner stress and exterior stress forms the current boundary surface configuration. Viewing from the exterior space of unit material, near the boundary surface, the scalar temperature gradient can be viewed as a constant. By this opinion, the unit boundary surface deformation is determined by quantum mechanics laws and the boundary surface is defined by a potential well. Near the potential well, the exact surface configuration is a quantum wave form. If too much detail is under examination, one will sink into quantum mechanics methods. However, that way will leads to statistic physics also. Hence, this paper will not go in quantum mechanics direction. Rather, in this research, the semi-empirical way is taken as following.

As this research is interesting on the macro thermo-effects, the surface motion will be taken as a quantum wave function form. The stress will defined on the exterior surface of unit boundary as it is a convention in traditional continuum mechanics. So, the scalar temperature gradient can be viewed as a constant.

For fixed scalar temperature gradient, the equation is simplified as:

$$\sigma_{ij} = \frac{\partial T}{\partial x^l} \frac{\partial G_i^l}{\partial x^j} \tag{16}$$

It shows that, as a first order approximation, the thermo stress is determined by the temperature scalar gradient and surface deformation. This stress is mainly related with thermo motion.

To clear this point, as a comparing, for conventional infinitesimal deformation, the unit surface deformation gradient is omitted. Furthermore, the surface transformation can be measured by the distance transformation, when the volume has no variation. Then, the elastic stress should be defined as:

$$\sigma_{ij}^{elas} = \frac{\partial^2 T}{\partial x^l \partial x^j}(G_i^l - \delta_i^l) \approx \frac{\partial^2 T}{\partial x^l \partial x^j}(F_l^i - \delta_l^i) \approx \frac{\partial^2 T}{\partial x^l \partial x^j}\varepsilon_{il} \tag{17-1}$$

Where, the $\varepsilon_{ij} = \frac{1}{2}[(F_j^i - \delta_j^i) + (F_i^j - \delta_i^j)]$ is the classical elastic strain.

For simple idea isotropic elastic continuum, the elastic stress is:

$$\sigma_{ij}^{elas} = \frac{\partial^2 T}{\partial x^l \partial x^j}(F_l^i - \delta_l^i) = \lambda(F_l^l - 3)\delta_j^i + 2\mu(F_j^i - \delta_j^i) \tag{17-2}$$

It implies that:

$$\frac{\partial^2 T}{\partial x^l \partial x^j} = \begin{cases} 2\mu \delta_{lj}, l \neq j \\ \lambda + \frac{2}{3}\mu, l = j \end{cases} \tag{17-3}$$

Where, repeat index summing is not used.

As the internal temperature variation (energy distribution) is determined by the intrinsic structure of continuum, the elastic parameters ($\lambda$ and $\mu$) are taken as the material feature parameters (constants). Surely, they are temperature dependent. From inverse-solution point of view, this equation will lead to lattice model. For a detailed theoretic treatment of elastic constants by the two order potential field of basic unit of materials, please referrer Born and Huang (1956, [12]). Where the zero potential gradient state is taken as the reference configuration, the order two differential of potential is used to obtain the elastic constants based on quantum wave function as a solution of potential well caused units-distance vibration.

Hence, the thermo stress first order approximation is well expressed by equation (16) while the elastic stress is well expressed by equation ((17-1). As a logic conclusion, the first item of equation (15) relates the coupling of micro unit thermo motion and macro deformation directly. Hence, the general stress definition equation (15) is justified in form.

Based on above formulation, the conclusion is that: 1) For large surface area variation (which contains the material under consideration), the stress is a rank two covariant tensor; 2) For purely the sequence order related energy transportation variation, the infinitesimal deformation, the stress is a rank two mixed tensor, although in intrinsic sense it is a rank two covariant tensor; 3) For general motion and deformation, the energy transportation gradient corresponds to temperature caused stress, the variation of energy transportation gradient corresponds to deformation stress. Both stresses are coupled by the deformation tensor.

### 3.3 Thermo Stress

Specially, for purely local rotation, the thermo stress is determined by the rotation. Generally, this is a harmonic vibration under potential well. Therefore, the thermo stress is expressed by the frequency and amplitude of vibration of quantum wave function. For potential gradient field, the unit surface motion is quantum motion.

To under stand the entropy concept, the most essential thermo motion in solid has no volume variation defined by $g^{ij} = g_0^{ij}$ and $g_{ij} = g_{ij}^0$. For volume invariant case $F_j^i = R_j^i$, a local rotation in continuum is expressed by the unit orthogonal tensor defined as:

$$G_i^j = R_i^j = \delta_i^j + \sin\Theta \cdot L_i^j + (1-\cos\Theta) L_i^l L_l^j \tag{18}$$

Where, $L_i^j = -L_j^i$ is the rotation direction tenser, $\Theta$ is local rotation angle. Generally, introducing rotation direction unit vector $L_i$, the rotation direction tenser components are: $L_2^1 = L_3$, $L_3^2 = L_1$, $L_1^3 = L_2$.

Typically, if the local rotation direction is fixed, one has thermo stress:

$$\sigma_{ij} = \frac{\partial T}{\partial x^l} \frac{\partial G_i^l}{\partial x^j} = \frac{\partial T}{\partial x^l} \cdot (\cos\Theta \cdot L_i^l + \sin\Theta \cdot L_i^k L_k^l) \cdot \frac{\partial \Theta}{\partial x^j} \tag{19}$$

This is the force supply the needed action to drive the heat transportation. Such a kind of thermo system is the most common one, and, as the first order approximation, it is the simplest one.

For periodic structure, as the quantum motion is typical (omitting initial phase), one has:

$$\Theta = \Theta_0 \cdot \sin[(p_l x^l - \omega t)] \tag{20-1}$$

$$\frac{\partial \Theta}{\partial x^j} = p_j \cdot \Theta_0 \cdot \cos[(p_l x^l - \omega t)] \tag{20-2}$$

Where, $p_l$ is wave number, $\omega$ is angular frequency.

As a first order infinitesimal local rotation angle approximation, one has:

$$\sigma_{ij}^\Theta \approx \frac{1}{2} p_j \frac{\partial T}{\partial x^l} \cdot (L_i L_l - \delta_{il}) \cdot \Theta_0^2 \sin[2(p_m x^m - \omega t)] \tag{21}$$

Perpendicular to the quantum wave direction, the stress component is zero. So, this thermo stress is a plane stress on the quantum wave transportation direction normal plane. As the quantum wave number direction is determined by the potential well structure, the stress is quantum solution dependent.

Hence, the wave number plays an important role. The thermo stress is vibration form. Vibration frequency and wave number is doubled quantum wave quantity. Its amplitude is:

$$\hat{\sigma}_{ij}^{\Theta} \approx p_j \frac{\partial T}{\partial x^l} \cdot (L_i L_l - \delta_{il}) \cdot \Theta_0^2 \qquad (22\text{-}1)$$

This stress is supplied by the electromagnetic field among molecules. That is the so-called Maxwell-Lorenz stress. For isotropic case ($L_i = \sqrt{\frac{1}{3}}$, $p_i = p_0$, $\frac{\partial T}{\partial x^j} = q_0$) on statistical scale, the stress is a pressure:

$$\hat{\sigma}_{ij}^{\Theta} \approx p_j \frac{\partial T}{\partial x^l} \cdot (L_i L_l - \delta_{il}) \cdot \Theta_0^2 = -\frac{2}{3} p_0 q_0 \cdot \Theta_0^2 \delta_{ij} \qquad (22\text{-}2)$$

The continuum is fluid-like.

Typically, if the local rotation angle is fixed, one has thermo stress:

$$\sigma_{ij}^{L} = \frac{\partial T}{\partial x^l} \frac{\partial G_i^l}{\partial x^j} = \frac{\partial T}{\partial x^l} \cdot (\sin\Theta \cdot \frac{\partial L_i^l}{\partial x^j} + (1-\cos\Theta) \cdot \frac{\partial}{\partial x^j}(L_i^k L_k^l)) \qquad (23)$$

For quantum motion, in form:

$$L_i^j = \overline{L}_i^j \cdot \sin(p_l x^l - \omega t) \qquad (24\text{-}1)$$

$$\frac{\partial L_i^k}{\partial x^j} = p_j \overline{L}_i^k \cdot \cos(p_l x^l - \omega t) \qquad (24\text{-}2)$$

So, the quantum wave form stress is:

$$\sigma_{ij}^{L} = p_j \frac{\partial T}{\partial x^l} \cdot [\sin\Theta \cdot \overline{L}_i^l \sin(p_m x^m - \omega t) + (1-\cos\Theta) \cdot \overline{L}_i^k \overline{L}_k^l \sin 2(p_n x^n - \omega t)] \qquad (25\text{-}1)$$

For infinitesimal local rotation angle, its first order approximation is:

$$\sigma_{ij}^{L} = p_j \Theta \cdot \frac{\partial T}{\partial x^l} \cdot L_i^l = \Theta p_j \cdot \delta_{ikl}(L_k \frac{\partial T}{\partial x^l} - \frac{\partial T}{\partial x^k} L_l) \qquad (25\text{-}2)$$

Where, the $\delta_{ikl}$ is Kronecher sign about ordering.

This stress is local rotation directional dependent. It is a plane stress perpendicular with the quantum wave direction. So, it is Maxwell-Lorenz stress. It is wave number dependent. Hence, the wave number plays an important role. Its isotropic form in average scale is:

$$\sigma_{ij}^{L} = 0 \qquad (25\text{-}3)$$

As a first order approximation, both cases give out the form conclusion:

$$\sigma_{ij} = \Theta^{\alpha} \cdot p_j q_l M_{il} \qquad (26)$$

Where, $q_l = \frac{\partial T}{\partial x^l}$, the tensor $M_{il}$ is purely determined by the structure tensor of lattice dynamic motion orientation. That is to say, the tensor $M_{il}$ is purely a dynamic geometry feature which is determined by the difference between the current configuration and the initial reference configuration. The success of statistical physics is lying that it catches this feature while letting the $q_l$ and $p_j$ as the basic point energy quantity and kinetic moment (wave number). In fact, the two quantities is the main topic for quantum mechanics. Hence, they are treated as known quantities in this research without the trouble to introduce the related quantum motion equations.

### 3.4 First Order Approximation of Thermo-stress

As special cases, for isotropic local temperature gradient, $q_l = \overline{q} = const, l = 1,2,3$. This model is suitable for most ball-like material unit. So, it can be taken as a good for the first approximation in general cases.

For fixed rotation direction, the first order infinitesimal local rotation angle approximation is:

$$\hat{\sigma}_{ij}^{\Theta} = \Theta_0^2 \cdot p_j \bar{q}_l (L_i L_l - \delta_{il}) \tag{27-1}$$

For fixed local rotation angle, the first order infinitesimal local rotation angle approximation is:

$$\hat{\sigma}_{ij}^{L} = \Theta \cdot p_j \bar{q}_l \cdot \bar{L}_i^l = \Theta \cdot p_j \cdot \delta_{ikl} (\bar{L}_k \bar{q}_l - \bar{q}_k \bar{L}_l) \tag{27-2}$$

The stress is determined by the quantum wave number and local rotation parameters. It shows that the orientation effects are much stronger than the angle effects.

Generally speaking, the simple thermo motion can be approximated by the lattice vibration. The problem here is in what extent the above approximation is a good one? For ball-like isotropic case ($L_i = \sqrt{\frac{1}{3}}$, $p_j = p_0$, $\bar{q}_l = \bar{q}_0$, $i,j,l = 1,2,3$), the pressure wave amplitude form is obtained as:

$$\hat{\sigma}_{ij}^{\Theta} \approx -\frac{2}{3} \Theta_0^2 \cdot p_0 \bar{q}_0 \cdot \delta_{ij} \tag{27-3}$$

$$\hat{\sigma}_{ij}^{L} \approx 0 \tag{27-4}$$

It shows that, for isotropic motion case, the orientation effects can be omitted and the angle effects play the main role.

Observing the sign of them, for conventional continuums, the negative sign is related with fluid. Hence, the fluid unit material mainly is unit orientation variation motion. The positive sign is related with gas. Hence, the gas unit material mainly is surface rotation angle variation. The phase transition is the related with the motion model variation.

Generally speaking, these two simplest equations relate the pressure, wave number, temperature gradient, and lattice vibration amplitude. Such a kind of relation equations can be obtained by using Hamiltonian defined with displacement perturbation. By solving the motion equation of displacement field, the similar results should be obtained. For this method, please refer a good review written by Carruthers (1961, [14]). Where, the frequency parameter is used. As the frequency and wave number is related with velocity, while the velocity must be supposed to be sound velocity or electromagnetic wave velocity, the results are not directly comparable with the results in this paper. However, the general features are similar. Personally, the method of starting from Hamilton principle with the simple assumption of there is a displacement field is not an exact description of unit motion. As a comment sense, the molecular local rotation in quantum sense will not cause distance variation among unit centers. Further, the unit surface will not necessarily change the unit volume. Hence, the distance still may be fixed. As a comment sense, the practice of traditional elastic deformation mechanics shows that the displacement field cannot exactly describe the unit motion provided an artificial condition called displacement-compatibility equations is introduced. The rotation cannot be well expressed by the strain symmetry assumption. For the topic of exact description of deformation, Chen (1987, [13]) has given out a detailed reasoning.

Reviewing some current reports, the pressure dependence on the wave number linearly (as first approximation) are reported by Peter, et al.(2007, [15]). For shock wave, the stress is mainly caused by wave number related deformation (Wallace, 1982, [16]). If the wave number vector and temperature gradient vector have a definite relation, the wave number can be replaced by the temperature gradient. Then, the thermo stress can be constructed by the tensor product of temperature gradient with itself (R Dominguez & D Jou, 1995, [17]). If the stress is interpreted as the source of migration velocity, the migration velocity will have a linear relation with the temperature gradient (P Rosenblatt, and V K LaMer, 1946, [18]).

Using $\Delta\phi = 0$ to introduce potential well concept, based on reasoning that the potential gradient and the temperature gradient has identical implication, the local rotation is completely determined by the temperature gradient (if the potential gradient is interpreted as force, and taking the quantum wave number as a temperature related parameter. Refer Equ.(27)). As the local orientation change and angle change are related with plastic deformation in macro sense, the temperature gradient and the plastic deformation are directly related. As a first order approximation, the similar formulation with above equations can be established. This topic is treated by F Zwicky (1932, [19]). On the other hand, if the stress is changed by outside applied stress, the temperature gradient will be changed. Hence, the heat conductivity will has a linear relation with the applied macro deformation stress (Kunzler, J E, et al. 1966, [20]).

For the effects of temperature gradient as the direct source and temperature gradient related stress (or migration position) oscillation, there are many papers. However, before much more works have been done, the direct comparing is difficult. However, they ([21]-[30]) are still proposed as the good reference documentation.

### 3.5 Thermo Stress under Volume Variation Caused by Large Local Rotation

For many kinds of molecular, the chemical action is expressed by the molecular active range expansion as the additional electrons are connected into the molecular. In rational mechanics, this is named as large local rotation. Physically, for free local rotation action, the action boundary of the basic unit has expansion. In this case, as the basic length vector transformation is in the form $F_j^i = \frac{1}{\cos\theta}\tilde{R}_j^i$, by the equation (2-2), the surface basic gauge vector transformation will be: $G_j^i = V \cdot \cos\theta \cdot \tilde{R}_j^i$. Where, the $\theta$ is the local rotation angle of $\tilde{R}_j^i$. Its geometrical meaning is that: for volume space, the additional electrons will force the area be contracted a little to let the additional electrons be in the space or expanded a little as a natural result of boundary expansion. For detailed geometrical field formulation please refer Xiao (2005, [13]).

When the active chemical molecular motion model are used as:

$$G_j^i = V \cdot \cos\theta \cdot \tilde{R}_j^i \tag{28}$$

Where, the volume scale $V(=\frac{V}{V_0}), V_0 = 1$ is introduced. It is clear that the thermo stress will be more complicated in form. For simplicity, infinitesimal local rotation angle is studied. By equation (16), the general stress is:

$$\sigma_{ij} = \frac{\partial T}{\partial x^l} \frac{\partial G_i^l}{\partial x^j} = \frac{\partial T}{\partial x^l} \cdot \frac{\partial(V\cos\theta)}{\partial x^j} \cdot \tilde{R}_i^l + V\cos\theta \cdot \frac{\partial T}{\partial x^l} \cdot \frac{\partial \tilde{R}_i^l}{\partial x^j} \tag{29}$$

For isotropic homogeneous volume variation:

$$\sigma_{ij} = \frac{\partial T}{\partial x^l} \cdot \frac{\partial G_i^l}{\partial x^j} = V \cdot \frac{\partial T}{\partial x^l} \cdot \frac{\partial (\cos\theta)}{\partial x^j} \cdot \tilde{R}_i^l + V\cos\theta \cdot \frac{\partial T}{\partial x^l} \cdot \frac{\partial \tilde{R}_i^l}{\partial x^j} \tag{30}$$

For fixed rotation direction, the first order infinitesimal local rotation angle approximation is:

$$\hat{\sigma}_{ij}^{\theta} \approx V \cdot \frac{\partial T}{\partial x^l} \cdot p_j \cdot \theta_0^2 \cdot [-\delta_i^l + \frac{1}{2}(\tilde{L}_i \tilde{L}_l - \delta_{il})] \tag{31-1}$$

Its isotropic stress amplitude form is:

$$\hat{\sigma}_{ij}^{\theta} \approx -\frac{4}{3} V \cdot q_0 \cdot p_0 \cdot \theta_0^2 \cdot \delta_{ij} \tag{31-2}$$

Comparing with equation (21), the volume variation related stress item is:

$$\hat{\sigma}_{ij}^{\theta V} = -V \cdot \frac{\partial T}{\partial x^i} \cdot p_j \cdot \theta_0^2 \tag{32-1}$$

It shows that, the molecular action will significantly change the thermo stress, especially on the local rotation direction.

Its isotropic stress amplitude form is:

$$\hat{\sigma}_{ij}^{\theta V} = -V \cdot q_0 \cdot p_0 \cdot \theta_0^2 \cdot \delta_{ij} \tag{32-2}$$

For fixed local rotation angle, the first order infinitesimal local rotation angle approximation is:

$$\sigma_{ij}^{\tilde{L}} \approx V \cdot \theta \cdot \frac{\partial T}{\partial x^l} \cdot p_j \cdot \tilde{L}_i^l = V \cdot \theta \cdot p_j \cdot \delta_{ikl}(\tilde{L}_k \frac{\partial T}{\partial x^l} - \frac{\partial T}{\partial x^k} \tilde{L}_l) \tag{33-1}$$

The thermo stress is proportional to the volume (expansion or contraction) scale. This is a well known fact in thermo mechanics.

Its isotropic case, this stress is zero.

$$\hat{\sigma}_{ij}^{\tilde{L}} \approx 0 \tag{33-2}$$

Although for infinitesimal deformation the pressure is linearly proportional with volume, as the deformation is an implicated function of volume, the linear relation is effective in limited range.

The $G_j^i = R_j^i$ forms the first classification. The $G_j^i = V\cos\theta \cdot R_j^i$ forms the second classification.

Only for these cases, the strong dependence of thermo stress on temperature gradient and the quantum wave functions is exposed.

Note that, even for infinitesimal local rotation, the orientation direction variation is the first order effects, while the local rotation angle is the second order effects (Oswald & Dequidt, 2008, [31]-[32]. The volume variation effects are the second order effects. Therefore, the elastic-plastic deformation effects are coupled with the second order effects.

## 4. Isolated Equilibrium System

After above formulation for unit material, now, the continuum is viewed as a whole. The macro quantities in traditional thermo mechanics will be studied. The simplest case is the continuum has no energy input or output. For simple, it is referred as an isolated system.

For an isolated system without energy exchange with environment, in standard coordinator system, one has surface integration on the configuration:

$$\oiint_{\Sigma} \vec{q} \cdot d\vec{s} = \iiint_{\Omega} \nabla \cdot \vec{q} \cdot dV = 0 \tag{34-1}$$

That is:
$$\frac{\partial^2 T}{\partial x^j \partial x^j} = \sigma_{jj} = 0 \tag{34-2}$$

This is the traditional formulation for temperature distribution. When this condition is met, the system is named as thermo-equilibrium isolated system. When the temperature on the medium boundary is given as boundary conditions, the temperature over the whole continuum can be determined.

Then, in traditional continuum theory, the isolated system is interpreted by thermo stress as:
$$\sigma_{ll} = 0 \tag{35}$$

This is the rational definition for isolated system: traceless thermo stress system. This explains a long-lasting inquiry about why the incompressible medium assumption is so widely used in fluid dynamics and solid mechanics.

### 4.1 Solid or Fluid Continuum in Equilibrium State

For solid or fluid continuum in equilibrium state, the system has no volume variation. Both the surface deformation and distance deformation are infinitesimal small. The thermo motion is approximated by the infinitesimal local rotation without volume variation.

By equations (22) and (25), the equation (38) means that for an isolated system:

$$(p_j L_j) \cdot (q_l L_l) - (p_l q_l) = 0 \text{, for fixed local rotation direction} \tag{36-1}$$

$$p_1(L_2 q_3 - q_2 L_3) + p_2(L_3 q_1 - q_3 L_1) + p_3(L_1 q_2 - q_1 L_2) = 0 \text{, for fixed rotation angle} \tag{36-2}$$

Geometrically, using $\alpha_{Lp}$ and $\alpha_{Lq}$ to represent the angle between the local rotation direction and the wave number direction and temperature direction respectively, and using $\alpha_{pq}$ to represent the angle between the temperature gradient and the quantum wave direction, the former can be written as:

$$\cos\alpha_{Lp} \cdot \cos\alpha_{Lq} = \cos\alpha_{pq} \text{, for fixed rotation direction} \tag{37-1}$$

As a special case, this feature can be explained by the parallel of rotation direction with the temperature gradient or the quantum wave direction. This is P-wave mode (Born and Huang, 1956, [12]).

Geometrically, The latter can be rewritten as:

$$\alpha_{Lp} = \alpha_{pq} = \frac{\pi}{2} \text{, for fixed rotation angle} \tag{37-2}$$

The later can be explained as the orthogonal feature of local rotation direction with the temperature gradient and the quantum wave direction. That is the S-wave model (Born and Huang, 1956, [12]).

### 4.2 Gas or Fluid Continuum in Equilibrium State

For gas continuum in equilibrium state, the system has no volume variation. However, its distance deformation is $F_j^i = \frac{1}{\cos\theta}\tilde{R}_j^i$. This can be explained by the large coherent distance in statistic physics viewpoint. Its surface deformation is $G_j^i = \cos\theta \cdot \tilde{R}_j^i$. This can be explained by the hitting cross section decrease. The volume is invariant. For this case, by equations (31) and (33), the traceless thermo stress is expressed as:

$$(p_j \tilde{L}_j) \cdot (q_l \tilde{L}_l) - 3(p_l q_l) = 0 \text{, for fixed rotation direction} \quad (38\text{-}1)$$

$$p_1(\tilde{L}_2 q_3 - q_2 \tilde{L}_3) + p_2(\tilde{L}_3 q_1 - q_3 \tilde{L}_1) + p_3(\tilde{L}_1 q_2 - q_1 \tilde{L}_2) = 0 \text{, for fixed rotation angle} \quad (38\text{-}2)$$

Geometrically, using the similar procedure as the last section, the above equations can be written as:

$$\cos \alpha_{\tilde{L}p} \cdot \cos \alpha_{\tilde{L}q} = 3 \cdot \cos \alpha_{pq} \text{, for fixed rotation direction} \quad (39\text{-}1)$$

$$\alpha_{\tilde{L}p} = \alpha_{pq} = \frac{\pi}{2} \text{, for fixed rotation angle} \quad (39\text{-}2)$$

Therefore, the gas or fluid continuum is different with solid in P-wave model while the S-wave model is similar.

This mechanism explains why the gas and fluid has no fixed configuration while the solid has fixed configuration.

As a special solution for both cases, there is a statistical solution:

$$q_i = q_0, \quad p_i = p_0, \quad \tilde{L}_i = \frac{1}{\sqrt{3}} \quad (40)$$

The local rotation direction is fully random distribution, and the temperature gradient and the quantum wave direction are isotropic distribution. This model has been used in idea gas thermodynamics for a very long time.

**4.3 Equilibrium State in General Sense**

The above simple cases discussion shows that: if random distribution concept is applied for the equilibrium system, the micro deformation detailed information will be lost.

Note the simple difference about isolated system and closed system:

Isolated system: $\sigma_{ll} = 0$ (41-1)

Closed system $\frac{\partial \sigma_{ij}}{\partial x^j} = 0$ (41-2)

Note that, as the isolated system is defined by: $\sigma_{jj} = \frac{\partial^2 T}{\partial x^l \partial x^j} G_j^l + \frac{\partial T}{\partial x^l} \frac{\partial G_j^l}{\partial x^j} = 0$, the boundary surface deformation can exist in a spatial harmonic form. This topic should be treated by quantum mechanics methods. That is not the topic for this paper.

Observing the general stress defined in equation (15): $\sigma_{ij} = \frac{\partial^2 T}{\partial x^l \partial x^j} G_i^l + \frac{\partial T}{\partial x^l} \frac{\partial G_i^l}{\partial x^j}$, only when $G_i^l \approx \delta_i^l$, one can obtain: $\sigma_{ij} = \frac{\partial^2 T}{\partial x^i \partial x^j}$ for no unit surface deformation. Reasoning from this point, the conventional thermo mechanics takes rigid unit as its basic material element. This simplification omits the unit surface deformation and, as a result, cannot establish the relation between deformation and thermo motion. This point is examined by a paper written by Rinaldi and Brenner (2002, [33]) under the question: 'Is the Maxwell stress tensor a physically objective Cauchy stress?'

Based on the reasoning formulated in this research, the Maxwell stress is directly related with item $\sigma_{ij} = \frac{\partial T}{\partial x^l} \frac{\partial G_i^l}{\partial x^j}$, while the Cauchy stress (deformation stress) is directly related with $\sigma_{ij}^{elas} = \frac{\partial^2 T}{\partial x^l \partial x^j}(G_i^l - \delta_i^l)$. This point will be made clear more under the following sections.

In this research, the item $\frac{\partial^2 T}{\partial x^l \partial x^j} G_i^l$ is viewed as macro configuration initial stress. So, in rational mechanics, the pure thermo stress is defined as: $\sigma_{ij} = \frac{\partial T}{\partial x^l} \frac{\partial G_i^l}{\partial x^j}$. The thermo stress is purely caused by the unit material surface geometrical motion field and the molecular scale potential (energy) gradient. By this definition, the condition $\frac{\partial^2 T}{\partial x^l \partial x^j} G_i^l \approx \frac{\partial^2 T}{\partial x^i \partial x^j} = const$ is implied for thermo equilibrium system (without macro deformation) as the initial stress of reference configuration, so a general formulation can be established for thermo mechanics.

## 5. General Formulation

As a general definition, the general thermo-stress is used to define the temperature variation vector variation caused by micro unit surface deformation:

$$\frac{\partial \vec{q}}{\partial x^j} = (\frac{\partial^2 T}{\partial x^l \partial x^j} G_i^l + \frac{\partial T}{\partial x^l} \frac{\partial G_i^l}{\partial x^j}) \vec{g}_0^i = \sigma_{ij} \vec{g}_0^i \tag{42}$$

It can be explained that the thermo stress is related with the macro temperature variation among molecular sequence. In micro sense, the thermo stress is explained by the material local rotation or deformation of geometrical configuration although the molecular sequence is fixed. This understanding leads to the following classification.

Firstly, the mechanics deformation caused temperature variation is examined. Then, the general treatment is carried out.

### 5.1. Mechanical Deformation:

Mechanical infinitesimal deformation of continuum is formulated as:

$$G_i^l = \delta_i^l + \varepsilon_i^l \approx \delta_i^l \tag{43}$$

In this case,

$$\begin{aligned}\sigma_{ij} - \sigma_{ij}^0 &= (\frac{\partial^2 T}{\partial x^l \partial x^j} G_i^l + \frac{\partial T}{\partial x^l} \frac{\partial G_i^l}{\partial x^j}) - (\frac{\partial^2 T}{\partial x^i \partial x^j}) \\ &= \frac{\partial^2 T}{\partial x^l \partial x^j}(G_i^l - \delta_i^l) \approx \frac{\partial^2 T}{\partial x^l \partial x^j} \varepsilon_i^l\end{aligned} \tag{44}$$

Where, $\sigma_{ij}^0 = (\frac{\partial^2 T}{\partial x^i \partial x^j})$.

For symmetrical stress, $\frac{\partial^2 T}{\partial x^l \partial x^j} \varepsilon_i^l = \frac{\partial^2 T}{\partial x^l \partial x^i} \varepsilon_j^l$, so:

$$\frac{\partial^2 T}{\partial x^l \partial x^j} \varepsilon_i^l + \frac{\partial^2 T}{\partial x^l \partial x^i} \varepsilon_j^l = 2\sigma_{il}^0 \varepsilon_j^l = 2\sigma_{jl}^0 \varepsilon_i^l \tag{45}$$

The universal form of elastic deformation stress is:

$$\sigma_{ij}^{elas} = \frac{1}{2}[\frac{\partial^2 T}{\partial x^l \partial x^j}\varepsilon_i^l + \frac{\partial^2 T}{\partial x^l \partial x^i}\varepsilon_j^l] = \frac{1}{2}[\sigma_{il}^0 \delta_{kl}\delta_{jm}\varepsilon_m^k + \sigma_{jl}^0 \delta_{kl}\delta_{im}\varepsilon_m^k] \qquad (46)$$
$$= C_{ijkm}\varepsilon_m^k$$

The tensor $C_{ijkm}$ is named as the elasticity constants tensor. That is:

$$C_{ijkm} = \frac{1}{2}[\sigma_{il}^0 \delta_{kl}\delta_{jm} + \sigma_{jl}^0 \delta_{kl}\delta_{im}] = \frac{1}{2}[\sigma_{ik}^0 \delta_{jm} + \sigma_{jk}^0 \delta_{im}] \qquad (47)$$

It is clear that it is temperature dependent by the definition or initial stress dependent.

Note that under the symmetry of strain assumption: $\varepsilon_m^k = \varepsilon_k^m$, one has:

$$C_{ijkm} = C_{jikm} = C_{jimk} = C_{ijmk} \qquad (48\text{-}1)$$

As the initial stress only has three independent component, combining with the condition that $\sigma_{ll}^0 = 0$, only two independent parameters are needed to defined the elasticity constants tensor.

Then:

$$C_{ijkm} = \lambda \delta_{ij}\delta_{mk} + 2\mu \delta_{jm}\delta_{ik} \qquad (48\text{-}2)$$

So, the elastic deformation constants depend on the initial intrinsic thermo-stress of material. As the initial thermo stress is determined by the statistical micro scale motion, hence, the lattice dynamics can give out a reasonable solution.

In many applications, the temperature dependence of elastic constant are expressed as:

$$\lambda = \lambda(T_0) + \frac{\partial \lambda(T_0)}{\partial T}\cdot \delta T, \quad \mu = \mu(T_0) + \frac{\partial \mu(T_0)}{\partial T}\cdot \delta T \qquad (49)$$

The formulation in this research is based on thermo mechanics. There are many text books try to make reader believe that the mechanical deformation can be interpreted by the thermo mechanics principles. Here, the rational forms are given. For stressed crystals, the elasticity problem is discussed under Green strain definition (Wallace, 1965, [34]).

The initial stress and conventional elasticity is related by:

$$C_{ijkm} = \frac{1}{2}[\sigma_{ik}^0 \delta_{jm} + \sigma_{jk}^0 \delta_{im}] = \lambda \delta_{ij}\delta_{mk} + 2\mu \delta_{jm}\delta_{ik} \qquad (50\text{-}1)$$

Generally speaking, this equation has no solution. Roughly, one has the approximation:

$$\lambda \approx \frac{1}{3}(\sigma_{ll}^0), \quad \mu = \frac{1}{6}(\sum_{i \neq j}\sigma_{ij}^0) \qquad (50\text{-}2)$$

For many solid materials in engineering, $\lambda \approx \mu$. For many fluids, $\lambda \approx 0$.

Analyzing the conventional deformation energy form $\Delta U = \sigma_{ij}\varepsilon_{ij}$ (Note that in some books, $\Delta U = \rho\sigma_{ij}\varepsilon_{ij}$, here $\rho$ is mass density. Based on deformation mechanics, the mass density should not appear here. This topic is arguable and will be discussed in the last section of this paper), as the strain is taken as the motion measurement, by the traditional entropy form definition $\Delta U = TdS$, a seasonable conclusion is that the entropy variation caused by mechanical deformation is proportional with the contraction of strain tensor with itself or the equivalent contraction of stress tensor with itself. By this meaning, as both are determined by the general configuration motion, the entropy represents the general motion measure in geometrical configuration. This topic is expanded in the next section.

### 5.2. Entropy: as a Measure of General Motion

As another limit of macro statistic approximation, in thermo dynamics initial macro effects can be introduced as an initial general stress (reference stress) defined under the condition of no surface deformation.

$$\sigma_{ij}^0 = \frac{\partial^2 T}{\partial x^l \partial x^j} \delta_i^l \tag{51}$$

Then, the measurable incremental stress can be introduced as::

$$\delta\sigma_{ij} = \sigma_{ij} - \sigma_{ij}^0 = \frac{\partial^2 T}{\partial x^l \partial x^j}(G_i^l - \delta_i^l) + \frac{\partial T}{\partial x^l}\frac{\partial G_i^l}{\partial x^j} \tag{52}$$

Where, the thermo stress variation is completely produced by the boundary surface variation containing the distinguishable material unit.

By this formulation, for average unit material, one has unit element net heat quantity variation as:

$$dQ = VdP \approx [\frac{\partial^2 T}{\partial x^l \partial x^j}(G_i^l - \delta_i^l) + \frac{\partial T}{\partial x^l}\frac{\partial G_i^l}{\partial x^j}] \cdot \frac{g^{ij}}{V} = T \cdot dS \tag{53}$$

Where, the unit material surface gauge tensor is $g^{ij} = \vec{g}^i \cdot \vec{g}^j$. Here, the geometrical equations $g^{il}g_{lj} = V^2 \delta_{ij}$, and $\vec{g}^i \cdot \vec{g}_j = V\delta_j^i$ are used. For details, please refer [11].

Comparing with above idealizations, the more appropriate entropy definition for unit surface deformation should be:

$$dS = \frac{1}{T}[\frac{\partial^2 T}{\partial x^l \partial x^j}(G_i^l - \delta_i^l) + \frac{\partial T}{\partial x^l}\frac{\partial G_i^l}{\partial x^j}] \cdot \frac{g^{ij}}{V} \tag{54}$$

Based on above formulation, the entropy should be defined as:

$$S = \frac{1}{T}\sigma_{ij}\frac{g^{ij}}{V} = \frac{1}{T}(\frac{\partial^2 T}{\partial x^l \partial x^j}G_i^l + \frac{\partial T}{\partial x^l}\frac{\partial G_i^l}{\partial x^j})\frac{g^{ij}}{V} \tag{55}$$

The initial entropy is defined by $G_j^i = \delta_j^i$ and $V_0 = 1$ as:

$$S^0 = \sigma_{ij}^0 \frac{g_0^{ij}}{V_0} = (\frac{\partial^2 T}{\partial x^l \partial x^j}\delta_i^l + \frac{\partial T}{\partial x^l}\frac{\partial \delta_i^l}{\partial x^j})\frac{g_0^{ij}}{V_0} = \frac{1}{T}\frac{\partial^2 T}{\partial x^l \partial x^l} \tag{56}$$

Physically, as $Q = \frac{\partial^2 T}{\partial x^l \partial x^l}$ is the unit volume heat source, the definition equation (56) is equivalent with traditional definition in traditional thermo mechanics.

The entropy depends on the average local rotation on a closed surface boundary. As this angle depends on the quantum field variation within the surface and their interaction with out field, hence, the entropy variation can be interpreted as the results of quantum field motion.

Summing up above results, for deterministic point of view, the entropy geometrical tensor can be introduced as:

$$s^l = \frac{\partial G_i^l}{\partial x^j} \cdot \frac{g^{ij}}{V} \tag{57}$$

$$s^{lj} = G_i^l \cdot \frac{g^{ij}}{V} \tag{58}$$

Where, the scalar temperature variation vector is viewed as a material features.

By this formulation, as the entropy geometrical tensor is anti-covariant tensors. They are equivalent with the surface quantum motion features, while the scalar entropy is mixed with the temperature field. Hence, geometrically, as they are related with spatial curvature, the entropy geometrical tensor is the configuration curvature of motion.

The macro scalar entropy which is consistent with conventional definition should be defined as:

$$S = \frac{1}{T}[\frac{\partial^2 T}{\partial x^l \partial x^j} G_i^l + \frac{\partial T}{\partial x^l} \frac{\partial G_i^l}{\partial x^j}] \cdot \frac{g^{ij}}{V} = \frac{1}{T}\frac{\partial^2 T}{\partial x^l \partial x^j} \cdot s^{lj} + \frac{1}{T}\frac{\partial T}{\partial x^l} \cdot s^l \tag{59}$$

Where, the surface deformation is represented by the entropy geometrical tensor, temperature is the energy distribution. So, this definition is intrinsic one.

By this definition, the traditional entropy definition form: $dQ = TdS$ is reserved.

Reasoning from physical consideration, disorder motion has the curvature feature. So, the disorder degree is proportional with the entropy. On the other hand, the small the selected scale (in sequence particle scale) is, the bigger the entropy is. These conclusions well explain the traditional concept for scalar entropy.

To see the benefits of above formulation, the following problems are addressed. For temperature gradient field, the mechanical deformation entropy increase can be defined as:

$$dS_{mecha} = \frac{1}{T}[\frac{\partial^2 T}{\partial x^l \partial x^j} \cdot G_i^l + \frac{\partial T}{\partial x^l} \frac{\partial G_i^l}{\partial x^j}] \cdot [\frac{g^{ij}}{V} - \frac{g_0^{ij}}{V_0}] = \frac{1}{T}\sigma_{ij} \cdot [Vg_{ij} - V_0 g_{ij}^0] \tag{60}$$

For infinitesimal deformation, taking $V_0 = 1$, the following approximation can be made:

$$Vg_{ij} - g_{ij}^0 \approx (1+\varepsilon_{ll})(\delta_{il}+\varepsilon_{il})(\delta_{lj}+\varepsilon_{lj}) - \delta_{ij} \approx \varepsilon_{ll}\delta_{ij} + 2\varepsilon_{ij} \tag{61}$$

So, the mechanical entropy variation is formulated as:

$$dS_{mecha} = \frac{1}{T}\sigma_{ij} \cdot (\varepsilon_{ll}\delta_{ij} + 2\varepsilon_{ij}) \tag{62-1}$$

For mechanical deformation caused entropy variation, the general stress is taken as constant tensor. In traditional mechanics, they are defined as material features.

Naturally, for fixed configuration, the general stress variation will produce the thermo entropy variation. It is defined as:

$$dS_{thermo} = \frac{1}{T}[\frac{\partial^2 T}{\partial x^l \partial x^j}(G_i^l - \delta_i^l) + \frac{\partial T}{\partial x^l}\frac{\partial G_i^l}{\partial x^j}] \cdot \frac{g^{ij}}{V} = \frac{1}{T}(\sigma_{ij} - \sigma_{ij}^0) \cdot \frac{g^{ij}}{V} \tag{62-2}$$

In fact, for infinitesimal thermo deformation under the same temperature, the entropy variation between two configurations is expressed as:

$$\begin{aligned}\Delta S &= \frac{1}{T}\sigma_{ij}\frac{g^{ij}}{V} - \frac{1}{T}\sigma_{ij}^0 \frac{g_0^{ij}}{V_0} \\ &= \frac{1}{2T}(\frac{g^{ij}}{V} + \frac{g_0^{ij}}{V_0})(\sigma_{ij} - \sigma_{ij}^0) + \frac{1}{2T}(\sigma_{ij} + \sigma_{ij}^0) \cdot (\frac{g^{ij}}{V} - \frac{g_0^{ij}}{V_0}) \\ &= dS_{thermo} + dS_{mecha}\end{aligned} \tag{63}$$

It shows that the thermo stress and mechanical stress are two main contributors of local entropy variation. On the other hand, the thermo variation and the macro mechanical variation are coupled together through the general entropy definition equation (55).

Comparing with the conventional deformation energy definition $\Delta U = \sigma_{ij}\varepsilon_{ij}$, the extra item in equation (62) $\Delta \widetilde{U} = \sigma_{ij}\varepsilon_{ll}\delta_{ij} = (\sigma_{jj})(\varepsilon_{ll})$ comes from the thermo heat source contribution. It is volume variation related deformation energy and is referred as volume variation effects in plastic deformation. In fact, the item $\Delta \widetilde{U} = \sigma_{ij}\varepsilon_{ll}\delta_{ij} = (\sigma_{jj})(\varepsilon_{ll})$ is defined as plastic deformation energy in conventional plastic deformation mechanics.

In gas-liquid transition, the accommodation or salvation in liquid phase will increase the basic unit volume. As the liquid stress is approximated by the isotropic pressure $-P_{liq}\delta_{ij}$, this item will be: $\Delta \widetilde{U} = \sigma_{ij}\varepsilon_{ll}\delta_{ij} = (\sigma_{jj})(\varepsilon_{ll}) \approx -3P_{liq} \cdot \Delta V$. Hence, for gas-liquid transition, the entropy drop will depend on the pressure and volume variation. As the volume variation $\varepsilon_{ll}$ can be evaluated by displacement field, so the efforts to explain the entropy drop by formulating the unit displacement field seems to be effective (Knox, C J H, and L F Philips, 1998,[35]). On the other hand, for phase transition, the item $\Delta U_{thermo} = (\sigma_{ij} - \sigma_{ij}^0)\frac{g^{ij}}{V} \approx 0$ can be approximated for a phase. Therefore, phase transition is mainly related with accommodation or salvation.

As an opposite example, the supercooled liquid has a negative volume variation. Then, the entropy excess also can be explained by the (displacement) vibration field (Johuri, 2001, [36]).

If taking the total entropy as a reference, the entropy drop related with accommodation or combining also can be interpreted by other methods (Iruduyam & Henchman, 2009, [37]).

Summing above main results, the entropy variation is caused by two parts. One part is thermo stress variation, another part is volume variation and elastic deformation. For gas-liquid phase transition, as the thermo stress is near constant, the entropy variation mainly depends on volume variation. For liquid-solid phase transition, when the volume variation is very small, the entropy variation mainly is caused by the thermo stress variation. The formulation in this research can be used to reexamine the related experimental data. This work may be done at later time when a suitable condition is available.

**5.3 Entropy for Infinitesimal Local Rotation**

For infinitesimal local rotation without volume variation, by equations (27) and (57)-(59), one has the following results.

For fixed rotation direction, the entropy based on the first order infinitesimal local rotation angle approximation is

$$dS^\Theta = \frac{1}{T}\hat{\sigma}_{ij}^\Theta \frac{g_0^{ij}}{V_0} = \frac{V_0}{T}\Theta_0^2 \cdot p_j\bar{q}_l(L_iL_l - \delta_{il})\delta_{ij} = \frac{V_0}{T}\Theta_0^2 \cdot p_j\bar{q}_l(L_jL_l - \delta_{jl}) \qquad (64)$$

For the fixed rotation direction $L_3 = 1$, its form is:

$$dS^\Theta = -\frac{V_0}{T}\Theta_0^2 \cdot (p_1\bar{q}_1 + p_2\bar{q}_2) \qquad (65)$$

Here, the negative sign show the phase shift. Note that, for quantum motion, the wave number $p_i$ and the quantum wave frequency $\omega$ is related by the equation: $p_ip_i = \omega/c$, here $c$ is light velocity.

For fixed local rotation angle, the first order infinitesimal local rotation angle approximation is:

$$dS^L = \frac{1}{T}\hat{\sigma}_{ij}^L \frac{g_0^{ij}}{V_0} = \frac{V_0}{T}\Theta \cdot p_j \bar{q}_l \cdot \bar{L}_i^l \delta_{ij} = \frac{V_0}{T}\Theta \cdot p_j \bar{q}_l \cdot \bar{L}_j^l \qquad (66)$$

For the average local rotation direction variation amplitude, $\bar{L}_3 = 1$, its form is:

$$dS^L = \frac{V_0}{T}\Theta \cdot (p_2 \bar{q}_1 - p_1 \bar{q}_2) \qquad (67)$$

Both cases give out the same conclusion that the entropy variation is related with the physical quantities on the normal plane of lattice orientation direction. For isotropic plane motion case, only the angular vibration contributes the entropy variation. For random distribution of lattice orientation, the average entropy variation will be: $dS^\Theta = -\frac{2}{3}\frac{V_0}{T}\Theta_0^2 \cdot p_0 \bar{q}_0$, and $dS^L = \sqrt{\frac{1}{3}\frac{V_0}{T}}\Theta \cdot p_0 \bar{q}_0$. Its wave number linear-relation is exact. As the wave number can be explained as the number of "states", then, in one explains the $\frac{\bar{q}_0}{T}$ as the probability density function of "state", the statistic methods can be examined by the formulation in this paper. However, the statistic methods failing to distinguish the different coefficients will cause serious trouble in explaining measured experimental data.

Similarly, when there is volume variation, for fixed rotation direction with volume variation, the entropy is:

$$dS^\theta = \frac{1}{T}\frac{\partial T}{\partial x^l} \cdot p_j \cdot \theta_0^2 \cdot [-\delta_i^l + \frac{1}{2}(\tilde{L}_i \tilde{L}_l - \delta_{il})]g^{ij} \approx \frac{V^2}{T}\frac{\partial T}{\partial x^l} \cdot p_i \cdot \theta_0^2 \cdot [-\delta_i^l + \frac{1}{2}(\tilde{L}_i \tilde{L}_l - \delta_{il})] \qquad (68)$$

For isotropic case, the entropy variation caused by local rotation angle is:

$$dS^\theta = -\frac{4}{3}\frac{V^2}{T}q_0 \cdot p_0 \cdot \theta_0^2 \qquad (69)$$

For fixed local rotation angle with volume variation, the entropy is:

$$dS^{\tilde{L}} = \frac{1}{T}\sigma_{ij}^{\tilde{L}} \frac{g^{ij}}{V} = \frac{1}{T}\frac{\partial T}{\partial x^l}\theta \cdot\cdot p_j \cdot \tilde{L}_i^l g^{ij} \approx \frac{V^2}{T}\frac{\partial T}{\partial x^l}\theta \cdot\cdot p_i \cdot \tilde{L}_i^l \qquad (70)$$

For isotropic case, the entropy variation caused by orientation vibration is:

$$dS^{\tilde{L}} = 0 \qquad (71)$$

Generally speaking, for infinitesimal deformation, the curvature of motion can be omitted, hence: $dS = 0$ is implied for equilibrium system. For solids, this is possible. Hence, for the solid matter, the static quantum-thermo stress is zero.

However, for fluid matter, the curvature of motion is significant. For statistical isotropic motion case, the pressure parameter represents the average motion curvature or says the entropy. Hence, the pressure is not zero. $\sigma_{ij} = -p\delta_{ij}$ For heat sink fluid matter, the temperature is inward direction. So, $dS > 0$. For heat radiating fluid matter, the temperature tends to outward direction. So, $dS < 0$. There is a zero entropy variation state $dS = 0$ as the transition point on macro average scale. In fact, $dS = 0$ is defined as reversible condition in traditional treatment. If it is explained by entropy definition in this research, it means that the average micro unit surface deformation has no variation.

Based on the above reasoning, the heat sinking matter has increased entropy. The heat supply matter has decreased entropy. Then, on the transition interface, the average entropy is zero. As the zero entropy means rectangular strait motion, hence, the transition surface is an average pure plane surface. After passing the transition interface, the motion curvature sign is reversed. So, the entropy variation is negative. That is: $dS \leq 0$. The final results are the sequence comes into inertial sequence without any inter relationship. The free particles sets are formed.

For $dS \leq 0$ system, the heat sink motion will produce positive stress (evaporation), while the heat radiating motion will produce negative stress (condensation). For $dS \geq 0$ system, the reverse is true. Therefore, the evaporation gas and condensation gas has opposite stress sign.

For a $dS \geq 0$ solid system consuming heat continuously, its entropy is increased firstly. When the entropy is big enough $S \geq S_{criti}$, it will become fluid or gas featured as $dS = 0$. Then, the $dS \leq 0$ system is formed. Evaporation will be produced. Finally, the system has gone away.

For a $dS \leq 0$ hot gas system radiating heat continuously, its entropy is decreased continuously. When the entropy is small enough $S \leq S_{criti}$, it will become fluid. The further radiating will cause the material become solid.

So, for a given matter, two parameters can be introduced. One is the maximum entropy of solid state $S_{solid}$. Another is the maximum entropy of liquid state: $S_{liquid}$.

Reasoning from above consideration, the vacuum is defined by $S = 0$. Therefore, for any real existing physical field, $S > 0$, the real physical space-time is curvature. This just is the basic principle of general relativity.

Following this line, the initial entropy should be treated as a curvature space-time frame. As a limit side, the quantum motion is related with big entropy system. So, the deterministic quantum motion corresponds to a very curvature space-time. This point catches the kernel of current physics.

Comparing with mechanical deformation, the thermo deformation is the difference of micro quantum deformations. Hence, the entropy, as the scale of deformation variation, is the scalar curvature of deforming micro configuration. When the curvature can be expressed by the deformation tensors, the traditional scalar entropy is defined. They form the main body of traditional thermo-dynamics. For quantum motion, the deformation tensor gradient (rank three tensor) is used to express the entropy. Unfortunately, the quantum mechanics try to establish the operation rule under different curvature features. So, the linkage to thermo dynamics and mechanical deformation are cut down. The only bridge which is left for us is the curvature space-time concept. The value of this research is that it exposed out the intrinsic relation for matter motion.

**5.4 Derive Coherent Concept in Statistical Physics**

According to the discussion about unit material infinitesimal deformation, omitting the macro mechanical deformation stress, (Equs.(27),(31),and (33)), the quantum motion related thermo stress can be expressed in form as:

$$\sigma_{ij}^{thermo} = A \cdot p_j q_l M_{il} \tag{72}$$

Where, $A = \Theta$ or $V\theta$ for fixed rotational angle and $M_{il}$ is symmetrical tensor; $A = \Theta^2$ or $V\theta^2$ for fixed rotation direction; the tensor $M_{il}$ anti-symmetrical. The geometrical tensor $M_{ij}$ is purely expressed by the quantum motion of unit material.

Without losing generality, by suitable coordinator selection, for the symmetry geometrical tensor, the quantum thermo stress can be written as a form:

$$\sigma_{ij}^{thermo} = A \cdot p_j q_l m_{(l)} \delta_{il} = A \cdot p_j q_i m_{(i)} \tag{73-1}$$

This stress is on the normal surface of temperature gradient direction.

For anti-symmetry geometrical tensor, the quantum thermo stress can be written as a form:

$$\tilde{\sigma}_{ij}^{thermo} = A \cdot p_j q_l \delta_{ilk} \tilde{m}_k = A \cdot p_j \delta_{ilk} q_l \tilde{m}_k \tag{73-2}$$

Where, $\delta_{ilk}$ is Kronecher sign. For most case, this item can be omitted.

When the temperature gradient and the wave number are stochastic processes, the equation (73-1) can be expressed as the correlation function as:

$$\sigma_{ij}^{thermo} = A \cdot [p_j q_i m_{(i)}] \tag{73-3}$$

Where, the square bracket expresses taking statistical correlation operation. As the temperature is related with the potential force direction and the wave number is related with the matter wave transportation direction, this item can be explained as the identity of moment density distribution. By this way, the thermo stress is determined by the coherence between the coherence density multiplied by the quantum state and temperature variation vector (as it is purely expressed by the wave number parameter $p_j$ and temperature gradient parameter $q_l$).

By this understanding, the quantum entropy can be defined as:

$$S_{quan} = \frac{1}{T} \frac{g^{ij}}{V} \sigma_{ij}^{thermo} = \frac{1}{T\bar{V}} \cdot A \cdot [g^{ij} \cdot p_j q_i m_{(i)}] \tag{74}$$

Where, the $\bar{V}$ ($\frac{g^{ij}}{V} = \bar{V} \cdot \delta_{ij}$) is a characteristic volume scale, as the $g^{ij}/V$ is also a stochastic process. Surely, this scale is related with the spatial correlation function.

By this formulation, the entropy is defined by the spatial correlation density function of quantum stats (defined by potential well feature $q_i$, quantum wave number $p_i$, and a geometrical factor $m_i$).

Based on above similarity analysis, the simplest way is to explain the thermo stress as the average kinetic moment of particles (such as the molecular hitting interpretation in the random motion model).

For complicated structure, the quantum thermo stress can be written as:

$$\sigma_{ij}^{quan} = \frac{\partial T}{\partial x^k} \Gamma_{ji}^k \tag{75}$$

For fixed potential well structure (or landscape), the geometrical tensor $\Gamma_{ij}^k$ should be explained by the features of complicated structure of material unit. By this formulation, the quantum entropy should be written as:

$$S_{quan} = \frac{1}{TV} \cdot \frac{\partial T}{\partial x^k} \cdot (\Gamma_{ij}^k g^{ij}) \tag{76}$$

By pure mathematic point of view, explaining the $(\Gamma_{ij}^k g^{ij}) = \Delta x^k$ as the general displacement in a curvature space equipped with link function $\Gamma_{ij}^k$, the quantum entropy can be explained by the curvature of quantum state space. Under this sense, the temperature gradient is explained as the general force. This way tends to connect the quantum motion thermo mechanics with the gauge field theory. All these things have happed in statistic physics.

The purpose of above discussion is to show that: the formulation in this research has no intrinsic contradict with already existing achievements represented by statistic physics.

**6. Entropy Variation in Continuum Thermomechanics**

In this formulation, there are two kinds of motion must be controlled by deterministic equations.

One is the geometrical motion: $G_j^i$. It puts its key motion on the orientation or configuration variation of basic materials as a whole. It is a macro physical equation. The motion is configuration dependent of basic material sequence.

**6.1 Entropy Variation under Macro Configuration Variation**

As the stress and geometrical gauge tensor are macro deterministic quantities, the deterministic macro entropy variation between two states can be expressed as:

$$\Delta S = \frac{1}{T}\sigma_{ij}\frac{g^{ij}}{V} - \frac{1}{T}\sigma_{ij}^0\frac{g_0^{ij}}{V_0}$$
$$= \frac{1}{2T}(\frac{g^{ij}}{V} + \frac{g_0^{ij}}{V_0})(\sigma_{ij} - \sigma_{ij}^0) + \frac{1}{2T}(\sigma_{ij} + \sigma_{ij}^0)(\frac{g^{ij}}{V} - \frac{g_0^{ij}}{V_0}) \qquad (77)$$
$$= dS_{thermo} + dS_{mecha}$$

Here, the entropy under initial configuration and the entropy under current configuration are compared. The temperature is viewed as a background field.

Observing the mechanical stress definition, the mechanical entropy variation must be zero or positive. That is:

$$dS_{mecha} = \frac{1}{2T}(\sigma_{ij} + \sigma_{ij}^0)(\frac{g^{ij}}{V} - \frac{g_0^{ij}}{V_0}) \geq 0 \qquad (78)$$

This condition for mechanical deformation is frequently discussed in many textbooks.

Mathematically, the initial thermal stress $\sigma_{ij}^0 = \frac{\partial^2 T}{\partial x^l \partial x^j}\delta_i^l$ should be positive for stable surface boundary defined by $\frac{\partial T}{\partial x^i}\vec{g}^i = 0$. If $\sigma_{ij}^0 = \frac{\partial^2 T}{\partial x^l \partial x^j}\delta_i^l$ is negative, the surface boundary is unstable for stable surface boundary defined by $\frac{\partial T}{\partial x^i}\vec{g}^i = 0$. By geometrical point of view, the stable surface corresponds to the valley of potential field landscapes while the unstable surface corresponds to the peak of potential field landscapes.

In many textbooks, the positive mechanical entropy variation condition is used to prove the positive feature of elastic parameters and viscosity parameter. Here, it is viewed as the natural result for stable surface boundary condition of unit material in continuum.

So, for boundary surface deformation around stable configuration, the thermal entropy variation should have the same sign with volume variation. This is only possible for stable solid, liquid, and gas continuum without internal chemical action. Therefore, having stable unit material configuration should be taken as the definition of continuum without internal chemical action. This unit material classification is different from the macro configuration stability. For macro configuration, the differences between solid and fluid (gas) is that the solid has macro stable configuration also, while the fluid (gas) macro configuration is unstable.

For boundary surface deformation around unstable configuration, the entropy variation has opposite sign with volume variation. This is the case for continuum with internal chemical action. In modern thermodynamics, the continuum with internal chemical action is treated by the self-organizing system. Hence, the self-organizing continuum has no stable unit material configuration. Of cause, it has no stable macro configuration.

For simple isotropic stress case (for solid, fluid or gas), the isotropic form of mechanical entropy variation is:

$$dS_{mecha} = \frac{p}{T} dV, \text{ for stable surface deformation} \tag{79-1}$$

$$dS_{mecha} = -\frac{\tilde{p}}{T} dV, \text{ for unstable surface deformation} \tag{79-2}$$

Where, the $p$ and $\tilde{p}$ are positive. The equation (79-1) is well-known in thermodynamics. However, in many textbooks or papers, they are explained by other considerations. The equation (79-2) is common in self-organizing system related with chemical action or biological growing. In scientific research circle, the negative entropy is a hot point. Many papers have been published.

In fact, the stable deformation is implied in traditional theoretic treatment. Where, the deformation process is defined as: $dS \geq 0$. This condition is named as the entropy increasing principle. Here, it is understood as the stable surface deformation condition. Only when the surface is stable, the reversible process concept can be used for $dS_{mecha} = 0$. On this sense, the irreversible concept in traditional thermomechanics only means that $dS_{mecha} > 0$.

For thermo entropy variation, the entropy increasing principle will require that:

$$dS_{thermo} = \frac{1}{2T}(\frac{g^{ij}}{V} + \frac{g_0^{ij}}{V_0})(\sigma_{ij} - \sigma_{ij}^0) \geq 0 \tag{80}$$

As the item $\frac{1}{2T}(\frac{g^{ij}}{V} + \frac{g_0^{ij}}{V_0})$ is always positive, the only interpretation is that the intrinsic stress variation is always positive. In statistic physics, as the stress is explained by the correlation function, the intrinsic stress increasing is explained by the increase of coherence. For simple isotropic stress case (for solid, fluid or gas), the isotropic form of thermo entropy variation is:

$$dS_{thermo} = \frac{V}{T} dp \tag{81}$$

This equation is well-known in thermodynamics.

So, the simplest macro entropy variation form is:

$$\Delta S = \frac{V}{T} dp + \frac{p}{T} dV, \text{ for stable surface deformation} \tag{82-1}$$

$$\Delta \tilde{S} = \frac{V}{T} d\tilde{p} - \frac{\tilde{p}}{T} dV, \text{ for unstable surface deformation} \tag{82-2}$$

By above formulation, if $dS = dS_{mecha} + dS_{thermo} \geq 0$ is applied to a continuum, a natural result is that the local self-organization phenomenon can be attributed to the mechanical deformation. In fact, this tendency has become a common practice in mechanical engineering.

The above formulation explains why the continuum with chemical action has no stable configuration. This is the dominant feature to distinguish the stable continuum and unstable continuum.

The equations also show that: stable continuum and unstable continuum can exist together without macro entropy variation. In fact, this is the most common facts in daily life.

For a special case, when $\Delta S = 0$, one has approximation:

$$\frac{1}{T}(\sigma_{ij} - \sigma_{ij}^0)\frac{g^{ij}}{V} + \frac{1}{T}\sigma_{ij} \cdot (\frac{g^{ij}}{V} - \frac{g_0^{ij}}{V_0}) = 0 \tag{83}$$

As the first item (mechanical entropy) is always positive, the second item (thermo entropy) must be negative. It shows that, the accumulated deformation of solid will finally make the continuum become other kind of material (chemical activity) if the total entropy is invariant. This feature is tightly related with fatigue-cracking or material damage phenomenon.

By defining the initial stress or residual stress to view this phenomenon, the accumulated deformation of solid will make the continuum has large initial stress and finally make the material lost its elasticity. This topic is treated by author in other papers.

Finally, for simple unstable continuum, the zero entropy variation means that:

$$\Delta \tilde{S} = \frac{V}{T}d\tilde{p} - \frac{\tilde{p}}{T}dV = 0 \tag{83}$$

Its form solution is:

$$\frac{V}{\tilde{p}} = \frac{dV}{d\tilde{p}} = const \tag{84}$$

For isolated unstable continuum, it means that: the chemical action or biology growing will keep the volume-pressure ratio be constant. This is a common phenomenon for pure chemical action, as the linear relationship between the pressure and volume. For a given initial volume and pressure, the constant is uniquely determined. However, any volume and pressure would be the solution also provided the volume-pressure ratio equal to the determined constant. This phenomenon is typical for hysteretic straining in solid (Bridgman, 1950, [38]). Where, positive stretching process produce a negative residual stress incremental, while the negative compressing process produce a positive residual stress incremental.

As a comparing, for simple stable continuum, the zero entropy variation means that:

$$\Delta S = \frac{V}{T}dp + \frac{p}{T}dV = 0 \tag{85}$$

Its form solution is:

$$pV = const \tag{86}$$

Once the constant is determined by the initial volume and pressure, the volume and pressure has an inverse relation.

### 6.2 Quantum Entropy and its Variation with Wave Number

For given temperature well structure, the quantum entropy is determined by the quantum motion, $p_i$. It puts its key motion on the physical state of basic materials. It is a micro physical equation. Based on previous formulation, the quantum vibration entropy amplitude can be defined as:

$$\hat{S}_{quan} = \frac{1}{T} \cdot \frac{\partial T}{\partial x^l} \frac{\partial G_i^l}{\partial x^j} \cdot \frac{g^{ij}}{V} = \frac{1}{T} \cdot \frac{\partial T}{\partial x^l} p_j \hat{G}_i^l \cdot \frac{g^{ij}}{V} \qquad (87)$$

Where, the $p_i$ is dominant wave number, the $\hat{G}_i^l$ is surface (configuration) vibration amplitude in statistic sense. Here, the quantum mechanics operation $\frac{\partial G_i^l}{\partial x^j} = \sqrt{-1} \cdot p_j G_i^l$ is used. For continuum. If the quantum mechanics operation is used to temperature field (the spatial periodic feature of units produce a spatial wave number $\tilde{p}_i$), then one will has the quantum entropy definition as: $S_{quan} = \tilde{p}_l p_j G_i^l \cdot \frac{g^{ij}}{V} > 0$. The quantum entropy is positive and is not zero. For isotropic ball-like particle model ($\hat{G}_i^l = \delta_i^l, \frac{g^{ij}}{V} = \delta^{ij}$), $S_{quan} = \tilde{p}_j p_j > 0$ the quantum entropy is completely determined by the spatial structure wave number and quantum matter wave number.

Generally speaking, the quantum state variation is caused by potential well variation. So, temperature variation is the main cause for quantum entropy variation. The reference quantum motion entropy can be defined by the initial temperature field and the related surface quantum motion as:

$$\hat{S}_{quan}^0 = \frac{1}{T_0} \cdot \frac{\partial T_0}{\partial x^l} p_{0j} \hat{G}_{0i}^l \cdot \frac{g_0^{ij}}{V_0} \qquad (88)$$

Therefore, the quantum motion change caused entropy variation is:

$$\Delta \hat{S}_{quan} = \frac{1}{T} \cdot \frac{\partial T}{\partial x^l} p_j \hat{G}_i^l \cdot \frac{g^{ij}}{V} - \frac{1}{T_0} \cdot \frac{\partial T_0}{\partial x^l} p_{0j} \hat{G}_{0i}^l \cdot \frac{g_0^{ij}}{V_0} \qquad (89)$$

By macro viewpoint, the quantum variation is mainly related with the wave number variation for infinitesimal thermo motion. Hence, for continuous measurement consideration, the macro quantum entropy amplitude variation can be approximated as:

$$d\hat{S}_{quan} = \frac{1}{T} \cdot \frac{\partial T}{\partial x^l} \hat{G}_i^l \cdot \frac{g^{ij}}{V} (p_j - p_{0j}) = (\frac{1}{T} \cdot \frac{\partial T}{\partial x^l} \hat{G}_i^l \cdot \frac{g^{ij}}{V}) \cdot dp_j \qquad (90)$$

It must be pointed out that: the surface quantum motion is related with the electron displacement motion (or velocity variation). As the Hamiltonian principle is used to derive the related motion equations, the wave number item will be mass density dependent. Generally speaking, the wave number is proportional with the mass density. If the mass parameter is introduced, the quantum entropy will be mass dependent linearly. However, for macro thermo entropy variation and mechanical entropy variation, the mass is implied by the temperature gradient function. Hence, no direct mass dependency appeared.

As the wave number is proportional with frequency, so the quantum entropy variation is proportional with frequency for incremental entropy variation. In fact, this feature is widely used in radioactive related technology. Here, a rational formulation and interpretation are given out.

### 6.3 Total Entropy Variation

When the temperature variation is taken into consideration, the macro configuration dependent entropy variation will be reformed as:

$$\Delta S = \frac{1}{T}\sigma_{ij}\frac{g^{ij}}{V} - \frac{1}{T_0}\sigma_{ij}^0\frac{g_0^{ij}}{V_0}$$

$$= \frac{1}{2}(\frac{1}{T}\sigma_{ij} - \frac{1}{T_0}\sigma_{ij}^0)(\frac{g^{ij}}{V} + \frac{g_0^{ij}}{V_0}) + \frac{1}{2}(\frac{1}{T}\sigma_{ij} + \frac{1}{T_0}\sigma_{ij}^0) \cdot (\frac{g^{ij}}{V} - \frac{g_0^{ij}}{V_0}) \quad (91)$$

$$= dS_{thermo} + dS_{mecha}$$

Finally, the total entropy variation is defined as:

$$\Delta S_{total} = \Delta S + \Delta \tilde{S}_{quan} = \Delta S_{mech} + \Delta S_{thermo} + \Delta \tilde{S}_{quan} \quad (92)$$

Where, $\Delta \tilde{S}_{quan}$ is a periodic vibration function. The micro motion and macro motion are combined by the entropy variation rather than the energy variation. Hence, the entropy concept is more intrinsic than the energy concept.

The direct sum of entropy describes the basic motion of matter as a continuum.

In logic sense, for a given thermo stress field: $\sigma_{ij}$, there are many possible interpretation. To find out the unique solution, whether geometrical motion is known or quantum motion is known.

For given thermo stress field and geometrical motion, the motion of basic material is inferred. This forms the topic for material damage, fatigue-cracking, material degeneration, and so on. To proof the correctness of these inferring, the nuclear physics, atom-molecular physics play their main roles.

For given thermo stress field and quantum motion, the possible configurations of materials as a whole are studied. This forms the basic topic for condensation physics. This branch is the main topic for physical-chemistry. The materials science should be put into this classification.

As a logic conclusion, for industry application and theoretic advancement, for given geometrical motion and quantum motion, constructing the theoretic formation becomes the main topic. So, this should be the main topic for a rational thermo mechanics.

As the thermo stress can be well measured by modern technology, it should be viewed as the known quantity.

On these sense, the statistical physics studies the quantum motion, introducing possible configuration motion to construct the measurable thermo stress. Therefore, in statistical physics, the configuration motion is experimental or empirical.

For physical chemistry, the configuration motions are under studding, the quantum motion is taken as experimental or empirical.

For biology research, the configuration motions can be measured well according to the given thermo stress field. Where, the pains taking problem is to guess the possible quantum motion, such as in biology evolution engineering..

**6.4 Background Entropy**

Even the mechanical deformation and thermo stress is zero, the background quantum entropy is not zero.

$$\hat{S}_{quan}^0 = \frac{1}{T_0} \cdot \frac{\partial T_0}{\partial x^l} p_{0j} \hat{G}_{0i}^l \cdot \frac{g_0^{ij}}{V_0} \quad (93)$$

Its non-zero condition is that: there is a potential well and matter does exist. This item is positive, as it is defined by the quantum wave amplitude.

For cosmic physics, rewritten it by its initial definition, the cosmic background entropy can be defined as:

$$S^0_{\cos mic} = \frac{1}{T_0} \cdot \frac{\partial T_0}{\partial x^l} \frac{\partial G_i^l}{\partial x^j} \cdot \frac{g_0^{ij}}{V_0} \tag{94}$$

Its non-zero condition is that: there is a potential field well and the space is curvature. The gravity field satisfies this condition. Hence, the entropy of gravity field can be determined by its gauge tensor formulated in general relativity, such as Schwarz gauge field or de-Sitter gauge field. For much more detailed discussion, please refer related documents.

## 7. Comparing with Related Formulations

Now, it is time to comparing the results in this paper with other formulation about thermomechanics in continuum. The focus will been put on the main contrast between both. Only by exposing the predictive ability of this new formulation, the theory can be taken as a reasonable one. However, the comparing is limited for simplicity.

### 7.1 Entropy Variation for Stable Continuum and Unstable Continuum

In this research, the stable continuum is defined by the unit material has stable surface deformation. If the unit surface deformation is unstable, the continuum is defined as unstable continuum (usually related with chemical reaction or biology growing). The classical thermodynamics is only applicable to stable continuum.

As a simple example, the phase transition usually is related with the unstable surface deformation. It is reasonable to predict that the classical entropy variation formulation can not be directly applied to phase transition. This is a fact. Although many papers published tried to solve the phase transition problem, however, the classical entropy variation (even the entropy definition) must be redefined.

In classical thermo dynamics, the equation: $dS = \frac{1}{T}(Vdp + pdV)$ are widely used and accepted as the one of basic equation. However, as it is stated in equation (82-1), it is only true for stable unit surface deformation. The stable continuum is composed by the units without chemical reaction. However, for the unstable continuum with chemical reaction, the surface boundary is changed by chemical action. By the formulation equation (82-2), the entropy variation is: $d\tilde{S} = \frac{1}{T}(Vd\tilde{p} - \tilde{p}dV)$. For volume invariant thermo motion, there is no difference between them. However, for pressure invariant thermo motion, the classical theory givens out the wrong entropy variation sign.

The negative entropy is frequently reported by experiments. However, as this is viewed as physically "not admissible", the related interpretation based on classical entropy formulation is very confusing.

In fact, the deformation caused chemical reaction is universally observed in geological formations and various rock samples. They are also widely observed in laboratory material engineering experiments. Failing to explain these facts is the main shortage for classical thermodynamics. This shortage seriously damages the application of thermodynamics in industry.

As an opposition view, the real problems are why, for most chemical action in mixtures continuum, the classical entropy variation formulation works well? This is an accident case. For mixture, as the chemical reaction happening, the stable unit and unstable unit are mixed together in the same space volume region, the observable entropy variation will be looks like this:

$$\Delta S = \frac{\alpha}{T}(Vdp + pdV) + \frac{1-\alpha}{T}(Vd\tilde{p} - \tilde{p}dV) \qquad (95)$$

Where, the $\alpha$ is the fraction of stable unit material, the $1-\alpha$ is the fraction of unstable unit material. For very low chemical reaction rate, at any time $1-\alpha \ll 1$. The chemical reaction has little effects. Therefore, it is difficult to identify it.

As the thermo environments are same for both, the pressure variation should be the same ($dp = d\tilde{p}$), the equation becomes:

$$\Delta S = \frac{V}{T}dp + \frac{dV}{T}[\alpha p - (1-\alpha)\tilde{p}] \qquad (96)$$

In classical formulation, for chemical reaction taking in fixed volume condition (such as in liquid state), the only problem is to interpret the pressure variation suitably. Therefore, no significant difficulty appears in low chemical reaction cases. In classical mechanics, the plasticity should be attributed to this category. (Bridgman 1950 [38], Levitas 1997[39])

Only for the fraction of unstable unit material is bigger enough, the chemical reaction effects will become apparent. However, at present stage, there is little data available from published paper.

Although by the entropy variation, it is difficult to find out the difference, the volume-pressure relationship for isolated system gives out an apparent contrast. For unstable continuum, $\frac{V}{\tilde{p}} = \frac{dV}{d\tilde{p}} = const$, for stable continuum, $pV = const$.

The above mentioned points should be taken as the experimental evidence for supporting the formulation in this paper.

**7.2 Unit Material Element Description**

In this research, physically, the unit material element surface boundary is described by zero temperature gradient (potential well peak or valley, they form a surface). This corresponds to the potential well concept. In fact, the mutual potential concept is used as the basic feature to define a continuum. This point also is the basic stone for the statistic theory of thermodynamics (Walter Noll, 1955, [40]). In Noll's paper, the potential field is used. As the temperature is more appropriate for continuum, the difference is only in from.

However, introducing mutual potential concept will take the distance of units as the basic quantity, this is not accepted here. The main reason is that, the quantum stable position is potential well dependent rather distance dependent. Hence, only the continuum behavior around the potential well should be the key factors. In fact, the potential landscapes research confirms the potential well concept.

In statistical physics, the temperature is related with distribution function. Hence, the temperature gradient can be viewed as the gradient distribution concept (or density gradient). Some researches are available along this line (A Onuki,2007, [41]).

Contrasting with the Onsager transport theory, the temperature gradient in this research defined by commoving dragging coordinator system is taken as the basic physical field. Combining with the geometrical deformation as the basic geometrical motion field, the geometrical field is taken as the basic observable thermal motion quantity. The temperature is taken as an universal quantity. From physical consideration, in abstract sense, the temperature gradient is related with the potential well which supplies the needed force to make the units being unified into a continuum. Hence, the temperature gradient is not related with other thermo quantities (such as thermo force) on basic quantity sense. The Onsager transport theory (based on state space) is completely irreverent with the formulation here. In fact, the general thermodynamics theory based on state space is not acceptable as there is no evidence to prove they are really independent quantities. In fact, the pressure is not independent from temperature and volume variation. The thermal field theory formulated (Sonnino, 2009, [42]) only has symbolic sense without any real geometrical sense. Such a kind field theory has misleading feature originated from pure mathematic formulism.

In fact, how to construct the basic unit material model in continuum is a very essential problem. As an approximation, volume variation can be taken as the basic geometrical motion representation. One resent example is given by Bouchbinder & Langer (2009, [43]). However, lacking the exact deformation representation makes the pressure definition and entropy formulation based on Holmholtz free energy seems baseless. Furthermore, the elastic deformation and plastic deformation are not well formulated.

In a resent paper by Noll (2010, [44]), a very abstract mathematic formulation is proposed. Where, the basic unit material is equipped with an initial configuration and the basic motion is represented by the deformation. Combing with the heat transportation vector, taking temperature as the result of heat transportation, an very compact mathematic formulation system is established. The difference with the research of this paper is that: the entropy definition. In this research, only temperature is taken as the basic physical quantity, there is no need to introduce too many concepts although they may be very helpful.

Based on physical reality exposed by modern physics, the only basic quantities of continuum are the potential wells and its structure (represented by its first and second derivative). Hence, introducing too many quantities may cause serious problems underlying the formulation.

### 7.3 Constitutive Equations

As a tradition, on macro point, for infinitesimal motion, the constitutive equations are widely used as a way to define the material features.

The simple elastic constitutive equations is given by equations (17) and more general form for arbitral temperature field is given by equations (46)-(52), where the elastic constants are defined by the mechanical initial stress defined by equation $\sigma_{ij}^0 = \frac{\partial^2 T}{\partial x^i \partial x^j}$. This way is taking the potential to replace the temperature field. It is based on the reference configuration. For the general form of deformation energy, there is no essential difference from Green formulation (Green & Zerna, 1954, [45], where the initial stress concept is implied. In this research, the mechanical initial stress is formulated. This initial stress only is applicable for macro deformation related stress variation.

In this research, another thermo initial stress is introduced. It is defined as thermo stress: $\sigma_{ij} = \frac{\partial T}{\partial x^l} \frac{\partial G_i^l}{\partial x^j}$. For macro static continuum, the initial stress may also be caused by the lattice vibration. The lattice dynamics theory developed by Born and Huang (1956, [12]) has shown that the elastic constants can be completely explained by the lattice vibration. By this thermo stress definition, the pressure of fluid and gas in their traditional constitutive equations are formulated. For infinitesimal local rotation, when the orientation is fixed (fixed local rotation direction), the pressure is negative. It corresponds to fluid. When the local rotation is caused by the unit-unit distance expansion, a positive pressure will appear. This corresponds to gas. Generally speaking, the gas is mainly related with the orientation large vibration. As the random orientation distribution will cause positive pressure, the idea gas model will give out positive pressure. For more general cases, the positive or negative pressure are used to distinguish the gas and fluid (depending on the quantum solution and the potential well direction). Hence, the gas is the most complicated material and usually should be in fluid-gas coexisting state.

On the other hand, when the general reference stress $\sigma_{ij}^0 = \frac{\partial^2 T_0}{\partial x^i \partial x^j} + \frac{\partial T_0}{\partial x^l} \frac{\partial G_i^l}{\partial x^j}$ is used to calculate the material constants, the viscosity parameters will be introduced through the thermo stress item. For traceless initial stress represents the thermal equilibrium state, the viscosity of fluid and gas are explained by the thermal initial stress. In a more broad sense, when the mechanical stress dominated elasticity is gradually changed into the thermo stress dominated elasticity for some materials, the process will be highly temperature dependent as the thermo stress variation is mainly quantum motion dependent. Some reports and theoretic research papers are available on this topic (Migliori, et al. 2006, [47]; Eckart, 1948, [48]; Kantor, 1989, [49]; Langer, 2008, [50]). For fluid motion, the phenomenon should be more significant for high temperature variation cases. However, this job is waiting to be done at a later time.

From physical consideration, as the thermal stress item $\sigma_{ij} = \frac{\partial T}{\partial x^l} \frac{\partial G_i^l}{\partial x^j}$ is mainly related with the potential well and its quantum solution, it can be solved by quantum mechanics methods. Or, it can be measured by the radioactive spectral measurement and analysis. In fact, most statistic physics is a long this line. The mechanical initial stress item $\sigma_{ij}^0 = \frac{\partial^2 T}{\partial x^i \partial x^j}$ can be measured by stress-strain curves or calculated by given unit potential structure. Hence, the general stress is a physical reality rather than symbolic form. In fact, this has become practical as the sensor technology and calculation technology are highly developed near resent 20 years.

The above features make the way to construct the constitutive equations more practical and more physical soundness.

**7.4 Entropy Additive not Energy Additive**

In many thermomechanics formulations, a common mistake is use the energy additive as basic equations. This is not acceptable in thermomechanics. As the temperature in thermomechanics is a vital parameter, only the entropy is additive quantity. This point is clear from the entropy definition: $dS = \frac{dQ}{T}$. The absolute temperature plays a role as an energy scale measure. This point is very basic in thermodynamics. Failing to identify this feature and directly using Hamilton quantities defined by energy to formulate the thermomechanics equations is not admissible in physics.

In this research, through the total entropy variation, the mechanical deformation and thermo deformation are coupled together.

The entropy concept defined in this research is applicable for classical elastic and plastic deformation, and is also applicable for classical thermodynamics. Its applicability to quantum mechanics and cosmic entropy problem show its universal features. Surely, this concept is arguable.

### 7.5 About Motion Equations

In fact, the motion equations are universal for continuum when the constitutive equations are given. So, these equations are not included in this paper. Many thermomechanics formulation takes too fast step to establish the thermo motion equations. However, if the thermo motion is not accurately formulated, such efforts are in vain. So, this research will not go further on this topic.

## 8. End Words

The formulation is summered in last section. The only waiting to be answer question is how correct is the formulation? I have no final answer for this question. The all positive evidences which is known for me have been used up when they are suitable for the reasoning stage. Here, repeating these contents is not practical. So, the paper will ended by the following words.

Why the science must take the experimental results as its final evidence? There are two reasons. 1) The abstract theoretic formulation is correct in its defined ranges. However, its related interpretation is not. The acceptability of interpretation cannot be taken as the final evidence. This means that the logic consistence must play its role. 2) The correct or intrinsic catching interpretation may fail to get good support from the abstract theoretic formulation. In this case, the correct formulation is under searching. The correctness of such o kind of searching must be checked by all related experiments for the universality of constructed abstract form. The formulated empirical equations are waiting to be reformulated.

The common mistakes in scientific circle which drag the advancement of science is the well-known theory equipped with abstract theoretic formulation and well-accepted interpretation. Such a kind of drag mainly originates from text book and dominant papers published. The theoretic formulation is correct in forms. However, its related interpretation is too artificial.

Another problem is failed to express the correct or intrinsic interpretation in an abstract formulation system under the guidance of experiments. The wrong mathematic methods play their roles in this field. This means that the empirical equations cannot be raised into an universal abstract form by themselves.

The first kind of science is referred as: basic theory. The second of science is referred as: phenomenon or empirical theory. The logic conclusion is that: to promote the advancement of science, the basic theory must be under upgrading of interpretation. The empirical equations must be under trying of unification in broad range. The interaction between the two domains will form the active force to drive the advancement of science as a whole, although the trace may be very curvature. I think the value of this research is lying here.